\documentclass[11pt]{article}
\usepackage{mydef2col}
\usepackage{myArticle}

\title{Semi-analytical pricing of American options with hybrid dividends via integral equations and the GIT method}
\author{
    \authorstyle{ Andrey Itkin}
    \newline \newline
    \institution{FRE Department, Tendon School of Engineering, New York University, email: \url{aitkin@nyu.edu}} \\
}

\date{\today}

\begin{document}

\maketitle

\lettrineabstract{This paper introduces a semi-analytical method for pricing American options on assets (stocks, ETFs) that pay discrete and/or continuous dividends. The problem is notoriously complex because discrete dividends create abrupt price drops and affect the optimal exercise timing, making traditional continuous-dividend models unsuitable. Our approach utilizes the Generalized Integral Transform (GIT) method introduced by the author and his co-authors in a number of papers, which transforms the pricing problem from a complex partial differential equation with a free boundary into an integral Volterra equation of the second or first kind. In this paper we illustrate this approach by considering a popular GBM model that accounts for discrete cash and proportional dividends using Dirac delta functions. By reframing the problem as an integral equation, we can sequentially solve for the option price and the early exercise boundary, effectively handling the discontinuities caused by the dividends. Our methodology provides a powerful alternative to standard numerical techniques like binomial trees or finite difference methods, which can struggle with the jump conditions of discrete dividends by losing accuracy or performance. Several examples demonstrate that the GIT method is highly accurate and computationally efficient, bypassing the need for extensive computational grids or complex backward induction steps.}

\vspace{0.5in}
\section{Introduction}

The valuation of American-style options is a nuanced and critically important challenge in quantitative finance. Unlike European options, American options can be exercised at any time before expiration, creating a complex optimal stopping problem. The holder must continuously decide whether to capture the option's immediate intrinsic value or retain it for potential future gains.

This challenge is profoundly complicated by discrete dividends. These lump-sum payments cause predictable drops in the underlying stock price, creating a powerful incentive for early exercise just before the ex-dividend date. Accurately determining the optimal exercise boundary is therefore paramount.

Pricing these instruments is more than an academic exercise; it is a vital task for traders, risk managers, and arbitrageurs. The standard Black-Scholes-Merton framework, which assumes a continuous dividend yield, is ill-suited for handling large, discrete dividends. This limitation has driven decades of research into more sophisticated models, which generally fall into two categories: those that modify the underlying stock price process and those that develop numerical techniques to solve the ensuing free boundary problem.

Due to the vast existing literature, we provide a focused survey organized by the underlying methodology for treating discrete dividends. This review traces the field's evolution from its analytical origins to current numerical techniques. We direct the reader to the cited references for a comprehensive bibliography.

\paragraph{\small Compound option approach \& Escrowed dividend models.} One of the earliest and most influential contributions was made by \cite{roll1977,geske1979,whaley1981} for pricing American calls. Their solution, often termed the Roll-Geske-Whaley (RGW) formula, treats an American call on a stock with a single discrete dividend as a compound option. The option to exercise just before the dividend is viewed as an option on another option expiring at the ex-date. While elegant and analytical, the RGW model is limited to a single dividend and only applies to calls, as early exercise of puts is optimal at other times. A common simplification, the Escrowed Dividend Models, adjusts the initial stock price by subtracting the present value of all future dividends. This model is simple to implement but is known to systematically misprice options, particularly those deep in- or out-of-the-money, as it fails to accurately capture the diffusion of the stochastic process around the known jump.

\paragraph{\small Lattice-Based Methods (Trees).} Lattice methods, particularly binomial and trinomial trees pioneered by \cite{CoxRossRubinstein:79} offer a more flexible numerical framework. The primary challenge is modifying the tree to account for the discrete dividend drop. Two main approaches have emerged:
\begin{itemize}
\item The "Lump-Sum" approach. The tree is built based on the stochastic process of the underlying stock, and at an ex-dividend date, the value at each node is reduced by the dividend amount. This simple method can lead to computational issues, as nodes may fall below zero for large dividends, violating the lognormal assumption.

\item The "Shift" approach suggested by \cite{bos2002, HHLewis2003}. The tree is constructed for the "stochastic" component of the stock price (i.e., the stock price minus the present value of future dividends). The known, deterministic dividend payments are then added back at each node. This approach avoids negative probabilities and is generally considered more robust and accurate, forming the basis for many modern implementations.

\end{itemize}

\paragraph{\small Partial differential equation and finite-difference methods.}
The PDE framework, rooted in the fundamental Black-Scholes differential equation, provides a powerful alternative. The problem becomes solving a PDE with a free boundary (the optimal exercise boundary) and a jump condition at each ex-dividend date. The value of the option immediately before the dividend date, $V(S^-(t_d), t_d)$, must be equal to the value immediately after, $V(S^+(t_d) - D, t_d)$, where $D$ is the dividend amount, and  $S^-(t_d), S^+(t_d)$ are stock values right before and after the dividend is paid. Numerical methods like finite differences are used to solve this system, often providing greater accuracy and convergence properties than trees for a given computational cost. In more detail, see \cite{wilmott1993,hirsa2013,chiarella2006numerical,ItkinBook,Andersen2025} and references therein.

\paragraph{\small Monte-Carlo methods.} Monte-Carlo simulation for American options relies on a Lest-Square method (LSMC) proposed in \cite{longstaff2001} and various more recent modifications. Handling discrete dividends in LSMC methods involves adjusting the stock price paths at ex-dates and has become a popular method for high-dimensional problems where trees and PDEs become impractical. Also, a recent development in \cite{Huo2025} proposes a hybrid method that improves the accuracy and stability of the standard LSMC algorithm (but with no discrete dividends). The core idea is to construct an ansatz for the continuation value using a 1D finite-difference PDE solution, which then serves as an effective control variate in the regression step. This approach bridges the gap between PDE and Monte Carlo methods, leveraging the strengths of both to reduce pricing error and improve robustness.

\paragraph{\small Recent developments.} A promising alternative for pricing American options is the integral equation method, derived from decomposition formulas by \cite{Peskir2005} (see also \cite{vellekoop2011integral,ItkinCF2025,AndersenLake2024} and references therein). This approach offers significant computational advantages: solving the resulting integral equations can be substantially more efficient than using standard finite-difference methods. Furthermore, for models with a Gaussian kernel such as geometric Brownian motion, the Fast Gauss Transform can be applied to compute the integral efficiently. This technique achieves linear complexity in the number of temporal nodes, even for multifactor models, enabling fast and scalable pricing.

The integral equation approach has been further extended to handle dividend-paying assets in \cite{vellekoop2011integral,AndersenLake2024}. For example, \citeauthor{vellekoop2011integral} generalize the decomposition formula to models with more general asset and cumulative dividend processes, uncovering new properties of the exercise boundary. However, a key challenge arises that computing the expectation of the early exercise premium (EEP) requires the transition density of the underlying process, which is generally unknown in closed form for many models, especially those that are time-inhomogeneous.

To address this, in this paper we propose several tractable alternatives for computing the option price, transition density, and early exercise boundary (EB). The latter can be efficiently determined using the GIT method, developed by the author and co-authors in a series of papers and comprehensively described in \cite{ItkinLiptonMuraveyBook}.

Note that a key practical decision is whether to model dividends as a fixed cash amount or a known proportion of the stock price. The fixed cash model is more common for mature companies but introduces the risk of negative prices. The proportional dividend model avoids this but may be less realistic for short-term, declared dividends. The choice significantly impacts the optimal exercise strategy and the resulting option price.

Despite this extensive literature, a consensus on a single optimal method remains elusive, as the choice is highly context-dependent. Key factors include the number and size of dividends, computational constraints, and the requirement to calibrate the model to market implied volatilities. For the purpose of generality, in this paper we consider a flexible hybrid dividend model that can simultaneously incorporate both discrete and continuous dividend representations or be reduced to either case.

The remainder of this paper is organized as follows. \Cref{model} outlines the mathematical framework for pricing American options with hybrid dividends. We review the decomposition formula of \cite{Peskir2005} and derive an integral equation for the American option price. In \cref{sec:expectation}, we compute the European option price and transition density for a time-dependent geometric Brownian motion model with hybrid dividends. \Cref{secEBPut} derives a Volterra integral equation for the EB using the GIT method. We provide a rigorous treatment of the weak singularities inherent in such equations, including the known singularity in the EB's derivative at expiration. A change of variables is proposed to remove these singularities, ensuring robust numerical solutions. We also demonstrate that in the dividend-free limit, our equation reproduces well-known properties of the EB for American Calls and Puts. \Cref{sect_deAmer} discusses de-Americanization within our framework - the process of converting an American option price into an equivalent European option price with the same implied volatility. This transformation simplifies pricing and analysis, particularly when inferring local volatility from market prices of American options. We also explore an alternative approach based on the implied strike concept introduced in \cite{Skabelin2015}, and show that it offers greater computational efficiency. \Cref{numExp} presents numerical experiments comparing the performance of our method with the binomial tree across various scenarios. The final section concludes by discussing implications for both academics and practitioners and suggests directions for future research.

\section{Model} \label{model}

In this section, we present our approach to pricing American options with discrete dividends using a specific model for the underlying stock price. Although the proposed method is general and applicable to various models, we focus here on the time-inhomogeneous Geometric Brownian Motion (GBM) model widely used by practitioners. We choose this model for clarity and transparency, while other models could be treated in a similar way. To make the model more realistic, we incorporate different types of discrete dividends (compare, e.g., with \cite{vellekoop2011integral}). Specifically, we assume the stock price $S_t$ under a risk-neutral measure $\mathbb{Q}$ follows the stochastic differential equation (SDE):
\begin{align} \label{GBM-41}
d S_t &= [a(t) S_t - b(t)] dt + \sigma(t) S_t dW_t,
\end{align}
with
\begin{align} \label{divDef}
a(t) &= r(t) - q(t) - \sum_{i=1}^{N_p} d_i \delta(t - t_{i}), \quad d_i \in [0,1) \\
b(t) &= \sum_{j=1}^{N_d} D_j \delta(t - T_{j}), \quad
S\Big|_{t=0} = S, \quad (t,S) \in [0,T]\times [0,\infty). \nonumber
\end{align}
Here, $r(t) \in \mathcal{C}^1$ is the instantaneous interest rate, $q(t) \in \mathcal{C}^1$ is the borrow cost (which could also represent the continuous dividend yield if the model is used in such settings),
$\sigma(t) > 0$ is the volatility, which  is a continuous function of $t$,
$d_i, i \in [1,N_p]$ is the value of the proportional discrete dividend paid at time $t_i$ (the dividend payment is proportional to $S_{t_i}$ and released on discrete dividend dates), $N_p$ is the number of proportional dividend payments, $D_j, j \in [1,N_d]$ is the value of the discrete cash dividend paid at time $T_j$, $N_d$ is the number of cash dividend payments and $\delta(x)$ is the Dirac delta function \cite{as64}. The representation of discrete dividends via the Dirac delta function has been widely utilized in the existing literature; see, for example, \cite{HHLewis2003,Ballester2008,GuoChang2020} among others.

In \eqref{divDef} we assume that $S_t \ge 0 \,\, \forall t \in [0,T]$ where $T$ denotes the maturity of the American option under consideration. However, one must bear in mind that if the borrow cost and discrete dividends are high, the model might become mean-reverting with mean-reversion level $\theta(t) = b(t)/a(t) < 0$ and mean-reversion speed $\kappa(t) = -a(t) > 0$. In this case,
\begin{equation}
\lim_{t \to \tau} \EQ[S_t] =  S e^{-\int_0^\tau \kappa(k) dk} + \int_0^\tau e^{-\int_k^\tau \kappa(\chi) d\chi}  \kappa(k) \theta(k) dk.
\end{equation}
where $\EQ[\cdot]$ is the expectation under the risk-neutral measure $\mathbb{Q}$. Since $\theta(k) < 0$, the expected value $\EQ[S_\tau]$ could become negative at some time $0 < \tau \le \tau(0)$.

To resolve this issue, we impose an additional regularization condition
\begin{equation}
\forall \tau \in [0,\tau(0)]: S e^{-\int_0^\tau \kappa(k) dk} + \int_0^\tau e^{-\int_k^\tau \kappa(\chi) d\chi}  \kappa(k) \theta(k) dk > 0.
\end{equation}
This condition introduces further constraints on the borrow cost and proportional discrete dividends that the model can accommodate.

Strictly speaking, we must also impose the condition $0 \le D_j < S^-_{T_{j}}$ for all $j \in [1,N_d]$. This ensures that the stock price $S^+_{T_j}$ remains positive after each cash dividend payment. However, enforcing this restriction makes the cash dividend model state-dependent. As noted in \cite{AndersenLake2024}, dividends in practice typically satisfy this condition, so it typically does not matter unless the model parameters are extremely unusual. For this reason, we relax this requirement in our subsequent analysis and instead assume that admissible values of $D_j$ obey this constraint by default.

\subsection{The decomposition formula}

To price American option written on the underlying stock $S_t$ with the dynamics given by \cref{GBM-41,divDef}, we use the construction known as the {\it American option decomposition} and originated to \cite{Kim1990}, see, e.g., a short survey in \cite{ItkinCF2025} where this approach is further developed and generalized.

Let us first consider an American Put option and denote its price as $P(t,S)$. We distinguish the exercise ($\mathcal{E}$) and continuation ($\mathcal{C}$) regions as
\begin{align}
\mathcal{E} &= \Big\{ (u,S_u)\in[0,T)\times (l_x,\infty): \, P(u,S_u) = K - S_u \Big\} \\
\mathcal{C} &= \Big\{ (u,S_u)\in[0,T)\times (l_x,\infty): \, P(u,S_u) > K - S_u \Big\}, \nonumber
\end{align}
\noindent where $l_x$ is the left boundary of the $S_t$ domain which could be either $l_x = 0$ or $l_x = -\infty$ depending on the model. These two regions are separated by the EB $S_B(t)$, which is a time-dependent function of the time $t$. Note, that in this paper we don't consider the case of multiple EBs as the latter is discussed in detail in \cite{ItkinKitapbayev2025}.

Using the change of variable formula of \cite{Peskir2005}, the following proposition has been proved in \cite{ItkinKitapbayev2025}
\begin{proposition}[Proposition 1 in \cite{ItkinKitapbayev2025}] \label{prop1}
Conditional on $S_t = S$, the American Put price with a {\bf single} exercise boundary $S_B(t)$ can be represented by the following decomposition formula
\begin{align} \label{decompGen}
P \left(t, S \right) &= \EQ \left\{ \DF(t,T) (K - S_T)^+ | S_t = S\right\} + \int_t^T \DF(t,u) \EQ \left\{ \left[ r(u)(K-S_u) + \mu(u,S_u) \right] \mathbf{1}_{S_u \in \mathcal{E}} \right\} d u.
\end{align}
where $\DF(t,s) = e^{-\int_t^s r(u)\,du}$ is the deterministic discount factor, and $\mu(t,S)$ is the drift of the corresponding underlying process. Here, the first term represents the European Put option price $P_E\left(t, S \right)$ while the second term is the EEP which depends on the early exercise boundary $S_B(t)$.
\begin{proof}[{\bf Proof}]
See \cite{ItkinKitapbayev2025}.
\end{proof}
\end{proposition}

Using \cref{GBM-41,divDef}, the integrand in the right-hands part of \eqref{decompGen} can be represented as
\begin{align} \label{til}
r(u)(K-S_u) + \mu(u,S_u) &= r(u)(K-S_u) + a(u) S_u - b(u) = \ta(u)S_u + \tb(u), \\
\ta(u) &= a(u) - r(u) = - q(u) - \sum_{i=1}^{N_p} d_i \delta(u - t_{i}) < 0, \quad d_i \in [0,1), \nonumber \\
\tb(u) &= r(u) K - b(u) = r(u) K - \sum_{j=1}^{N_d} D_j \delta(u - T_{j}). \nonumber
\end{align}
Note, that $\tb(u)$ could be both positive and negative depending on values of $K, r(t), D_j$. Also, $\ta(t), \tb(t)$ are not continuous anymore and experience finite jumps at $t_{i}$ and $T_{j}$, respectively, see \cite{vellekoop2011integral,AndersenLake2024} among others.

Using the definitions in \eqref{til}, the decomposition formula can be re-written in the form
\begin{align} \label{decompGen2}
P \left(t, S \right) &= P_E \left(t, S \right) + \int_t^T \DF(t,u) \EQ \left\{ \left[ r(u)K - q(u) S_u \right] \mathbf{1}_{S_u \in \mathcal{E}} \right\} d u + \Theta(t,S), \\
\Theta(t,S) &= \sum_{j=1}^{N_d} \DF(t,T_{j}) D_j \int_0^{S_B(T_j)} G(S,\xi, T_j - t) d \xi
- \sum_{i=1}^{N_p} \DF(t,t_{i}) d_i \int_0^{S_B(t_i)} \xi G(S,\xi, t_i - t) d \xi, \nonumber
\end{align}
where $G(x,\xi,t)$ is the transition density function (in this case it coincides with Green's function of the corresponding forward PDE, see below).

In the absence of discrete dividends and with $q(t)$ denoting the continuous dividend yield, this formula appears in \cite{ItkinKitapbayev2025,ItkinCF2025}. For the time-homogeneous GBM model it has been presented in a number of papers, e.g., see \cite{Andersen2016,AndersenLake2021} and references therein and in \cite{ItkinCF2025}.

\section{Computing expectations in \eqref{decompGen}} \label{sec:expectation}

The decomposition formula Equation \eqref{decompGen2} provides a solution for pricing American options with discrete dividends, contingent upon knowing the early exercise boundary $S_B(t)$ and the density function $G(x, \xi, t)$ in either closed form or through numerical methods. However, for the model described in \cref{GBM-41,divDef}, these quantities are not readily available. In this section, we present a method to compute the European Put option price and the density of the process.

\subsection{The European option price} \label{Europ}

Even for the European options written on the stock $S_t$ with the dynamics provided in \cref{GBM-41,divDef} the explicit expression for $P_E(t,S)$ is not known. To proceed, observe that by a standard argument, \cite{ContVolchkova2005, klebaner2005} the European Put option price $P_E(t,S)$ solves a parabolic partial differential equation (PDE)
\begin{equation} \label{PDE}
\fp{P_E}{t} + \dfrac{1}{2}\sigma^2(t) S^2 \sop{P_E}{S} +  [a(t) S - b(t)] \fp{P_E}{S} = r(t) P_E,
\end{equation}
\noindent which should be solved subject to the terminal condition at the option maturity $t=T$
\begin{equation} \label{tc0}
P_E(T,S_T) = (K - S_T)^+,
\end{equation}
\noindent and the boundary conditions
\begin{equation} \label{bc0}
P_E(t,S)\Big|_{S \uparrow \infty} = 0, \qquad P_E(t,0) = K e^{-\int_T^t r(k) dk} .
\end{equation}

By a change of variables
\begin{align} \label{tr1}
S \to \frac{K}{\alpha(t)}e^x, \quad P_E(t,x) &= W(t,x)e^{\int_T^t r(k) dk} + K e^{-S/K -\int_T^t r(k) dk}, \quad \alpha(t) = e^{\int_T^t \left[ \frac{1}{2} \sigma^2(k) - a(k)\right] dk}, \\
(t, x) &\in [0,T]\times \mathbb{R} \nonumber
\end{align}
the PDE in \eqref{PDE} can be transformed to
\begin{align} \label{PDE1}
\fp{W}{t} &+ \dfrac{1}{2}\sigma^2(t) \sop{W}{x} - e^{-x} \alpha(t) \frac{b(t)}{K} \fp{W}{x} + \dfrac{1}{2}\sigma^2(t)\Theta(t, x) = 0, \\
\Theta(t, x) &= K e^{-\frac{e^x}{\alpha(t)}-2 \int_T^t r(k) \, dk} \left\{ \frac{e^{2x}}{\alpha^2(t)}
+ \frac{2}{\sigma^2(t)} \left[ \frac{b(t)}{K} - 2 r(t) - e^x \frac{a(t)}{\alpha(t)}\right] \right\}, \nonumber \\
\lim_{x \to \infty} \Theta(t, x) &= 0, \qquad \lim_{x \to -\infty} \Theta(t, x) =2 K\frac{b(t)/K - 2 r(t)}{\sigma^2(t)} e^{-2 \int_T^t r(k) \, dk}. \nonumber
\end{align}
Making another transformation of the time
\begin{equation} \label{tau}
\tau = \frac{1}{2} \int_t^T \sigma^2(k) dk \ge 0,
\end{equation}
we arrive at another PDE for $W(\tau,x)$
\begin{align} \label{PDE2}
\fp{W}{\tau} &= \sop{W}{x}  - e^{-x} \gamma(\tau) \fp{W}{x} + \Theta(\tau,x), \qquad \gamma(\tau) = 2 \frac{\alpha(t(\tau)) b(t(\tau))}{K \sigma^2(t(\tau))}, \qquad t = t(\tau),
\end{align}
where the function $t(\tau)$ can be obtained by inverting \eqref{tau}. Accordingly, the boundary condition \eqref{bc0} and the terminal condition in \eqref{tc0} in new variables read\footnote{Note, that by \eqref{tr1} we have $\alpha(\tau)|_{\tau = 0} = 1$.}
\begin{equation} \label{bc1}
W(\tau,x)\Big|_{x \uparrow \infty} = 0, \qquad W(\tau,x)\Big|_{x \uparrow -\infty}  = 0,
\qquad W(0,x_T) = K \left[ \left(1 - e^{x_T} \right)^+ -  e^{-e^{-x_T}} \right].
\end{equation}

The PDE in \eqref{PDE2} does not admit a closed-form solution. However, it can be transformed into a linear Volterra integral equation of the second kind, which can then be solved numerically. To see this, consider \eqref{PDE2} without the last term on the right-hand side. In this case, the equation reduces to a standard heat equation, whose Green's function is well-known \cite{polyanin2008handbook}.
\begin{equation} \label{Green}
G(x,\xi,\tau) = \frac{1}{2\sqrt{\pi \tau}} e^{- \frac{(x-\xi)^2}{4 \tau}}.
\end{equation}

Using this result, the additional term in \eqref{PDE2} can be taken into account by using a generalized Duhamel's principle, \cite{Itkin2024jd,Hunter2014}. This yields
\begin{align}
W(\tau, x) &= K \int_{-\infty}^{\infty} \left\{ \left( 1 - e^\xi \right)^+ - e^{-e^{\xi}} \right\} G(x, \xi, \tau) d\xi \\
&+ \int_0^\tau \int_{-\infty}^{\infty} \left[ \Theta(\nu, \xi) - e^{-\xi} \gamma(\nu) W'_\xi(\nu,\xi) \right] G(x, \xi, \tau-\nu) d \xi d \nu. \nonumber
\end{align}
Integrating the last term by parts yields
\begin{align} \label{Volt1}
W(\tau, x) &= K \int_{-\infty}^{\infty} \left\{ \left( 1 - e^\xi \right)^+ - e^{-e^{\xi}} \right\} G(x, \xi, \tau) d\xi + \int_0^\tau \int_{-\infty}^{\infty} \Theta(\nu,\xi) G(x, \xi, \tau-\nu) d\xi d\nu \\
&+ \frac{1}{2\sqrt{\pi}} \int_0^\tau \int_{-\infty}^{\infty} \frac{x - \xi - 2 (\tau - \nu)}{\tau - \nu} \frac{\alpha(\nu)}{\sigma^2(\nu)} b(\nu) e^{-\xi} G(x, \xi, \tau-\nu) W(\nu,\xi) d\xi d\nu. \nonumber
\end{align}

Due to the definition of $b(t)$ in \eqref{divDef}, the last double integral can be reduced to a single one. Therefore, \eqref{Volt1} transforms to a Volterra-Fredholm integral equation of the second kind
\begin{align} \label{Volt2}
W(\tau, x) &= K \int_{-\infty}^{\infty} \left\{ \left( 1 - e^\xi \right)^+ - e^{-e^{\xi}} \right\} G(x, \xi, \tau) d\xi + \int_0^\tau \int_{-\infty}^{\infty} \Theta(\nu,\xi) G(x, \xi, \tau-\nu) d\xi d\nu \\
&+ \sum_{j=1}^{N_d} D_j \bm{1}_{\tau_{j} \le \tau} \int_{-\infty}^{\infty} \frac{x - \xi - 2 (\tau - \tau_{j})}{\tau - \tau_{j}} \frac{\alpha(\tau_{j})}{\sigma^2(\tau_{j})} e^{-\xi} G(x, \xi, \tau-\tau_{j}) W(\tau_{j},\xi) d \xi. \nonumber
\end{align}
Here, discrete cash dividends are counted backward in time $t$ (or, equivalently, forward in time $\tau$). The notation $\bm{1}_{\tau_{j} \le \tau}$ indicates that we only consider dividends satisfying $0 < \tau_{1-} < \ldots < \tau_{k-} < \tau$, where $\tau_j$ is the inverse map of $T_j$ as defined in \eqref{tau}.

We have to solve \eqref{Volt2} by setting $t=0, \tau = t(0)$. Once it is solved, the European option price $P_E(t,x)$ follows from \eqref{tr1}.

The \eqref{Volt2} can be solved sequentially in time $\tau$ by first, setting $\tau = \tau_{1-}$, then $\tau = \tau_{2-}$, and so on. Observe, that at $\tau = \tau_{j}$ the following identity holds
\begin{equation}
\frac{x - \xi - 2 (\tau - \tau_{j})}{\tau - \tau_{j}} G(x, \xi, \tau-\tau_{j}) =
G'_\xi(x, \xi, \tau-\tau_{j}) - G(x, \xi, \tau-\tau_{j}) = \delta'(x - \xi) - \delta(x - \xi).
\end{equation}
Then, we proceed as follows:

\begin{myAlgorithm}{Pricing procedure} \label{proc}
\item $\bm{\tau < \tau_{1-}}$. The solution of \eqref{Volt2} coincides with the Black-Scholes formula with time-dependent coefficients.

\item $\bm{\tau = \tau_{1-}}$. Using the identity, \cite{as64}
\begin{equation} \label{deltaPrime}
\int_{-\infty }^{\infty } f(\xi) \delta '(x-\xi ) \, d\xi = f'(x)
\end{equation}
we obtain from \eqref{Volt2}
\begin{align} \label{unmodA}
W(\tau_{1-}, x) &= A(\tau_{1-}, x) + B(\tau_{1-}) e^{-x} \left[ W'_x(\tau_{1-},x) - 2 W(\tau_{1-},x)\right], \\
A(\tau,x) &= K \int_{-\infty}^{\infty} \left\{ \left( 1 - e^\xi\right)^+ - e^{-e^{\xi}} \right\} G(x, \xi, \tau) d\xi \nonumber \\
&+ \int_0^\tau \int_{-\infty}^{\infty} \Theta(\nu,\xi) G(x, \xi, \tau-\nu) d\xi d\nu, \qquad
B(\tau) = \frac{1}{2\sqrt{\pi}} D_1 \frac{\alpha(\tau)}{\sigma^2(\tau)}. \nonumber
\end{align}
This is a first-order ordinary differential equation for $W(\tau_{1-},x)$. By solving it, we obtain
\begin{equation}
W(\tau_{1-},x) = e^{\frac{e^x}{B(\tau_{1-})} + 2 x} \frac{1}{B(\tau_{1-})}\int_{x}^\infty A(\tau_{1-},k) e^{-k - e^{k}/B(\tau_{1-})} dk.
\end{equation}
At $x \to \pm \infty$ this yields $W(\tau_{1-},x) \to 0$ which are the correct boundary conditions, see \eqref{bc1}.

\item $\bm{\tau_{2-} > \tau > \tau_{1-}}$. Since $W(\tau_{1-},x)$ was computed in the previous step, the option value $W(\tau,x)$ is given by the Black-Scholes formula with time-dependent coefficients (the first term in \eqref{Volt2}) plus a second integral whose integrand is already known.

\item $\bm{\tau = \tau_{2-}}$. This step is similar to the step (b) with the only difference that $A(\tau_{2-},x)$ is now defined as
\begin{align} \label{modA}
 \hspace{-0.9\leftmargin}
 A(\tau_{2-},x) &= K \int_{-\infty}^{\infty} \left\{ \left( 1 - e^\xi \right)^+ - e^{-e^{\xi}} \right\} G(x, \xi, \tau_{2-}) d\xi + \int_0^\tau \int_{-\infty}^{\infty} \Theta(\nu,\xi) G(x, \xi, \tau_{2-}-\nu) d\xi d\nu \nonumber \\
&+ \frac{1}{2\sqrt{\pi}} \sum_{j=1}^{1} D_j \int_{-\infty}^{\infty} \frac{x - \xi + 2 (\tau_{2-} - \tau_{1-})}{\tau_{2-} - \tau_{1-}} \frac{\alpha(\tau_{2-})}{\sigma^2(\tau_{2-})} e^{-\xi} G(x, \xi, \tau_{2-}-\tau_{1-}) W(\tau_{1-},\xi) d \xi. \end{align}

\item and so on ...
\end{myAlgorithm}

\begin{figure}[!htp]
\centering
\fbox{\includegraphics[width=0.7\textwidth]{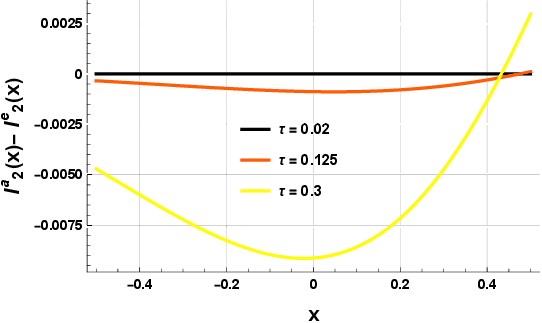}}
\caption{The difference between the "exact" (computed numerically) and approximating (by \eqref{appr1}) values of $I_2(x)$ for $\tau = 0.02, 0.125, 0.3$.}
\label{figApp1}
\end{figure}

It is important to notice, that the first integral in \eqref{modA} (and also in \eqref{unmodA}) could be computed in closed-from since the Green's function is Gaussian. This yields
\begin{align}
I_1 &= \int_{-\infty}^{\infty} \left(1 - e^\xi \right)^+  G(x, \xi, \tau) d\xi = \Phi\left(\frac{x + \log \left(B(\tau)\right)}{\sqrt{2\tau }}\right) \\
&+ e^{\tau +x} \Bigg[ - \Phi\left(\frac{\log \left(B(\tau)\right) + 2(\tau + x)}{\sqrt{2\tau }}\right) + \Phi \left(\frac{2 \tau +x}{\sqrt{2\tau }}\right) - \Phi \left(\sqrt{2(\tau+x) + \frac{x^2}{2\tau }} \right) + \frac{3}{2} \Bigg], \nonumber
\end{align}
where $\Phi(x)$ is the normal CDF, \cite{as64}. The integral
\begin{equation} \label{I2}
 I_2 = - \int_{-\infty}^{\infty} e^{-e^{\xi}} G(x, \xi, \tau_{2-}) d\xi
\end{equation}
can be approximated by using a Taylor series expansion of $e^{-e^x} - e^{-e^\xi}$ around $\xi = x$. For instance, taking into account the first four terms yields
\begin{align} \label{appr1}
I_2 &= \int_{-\infty}^{\infty} \left( - e^{-e^{\xi}} \right) G(x, \xi, \tau) d\xi = - e^{-e^{x}}  + \int_{-\infty}^{\infty} \left( e^{-e^{x}} - e^{-e^{\xi}} \right) G(x, \xi, \tau) d\xi \\
&\approx - e^{-e^{x}} + e^{x-e^x} \int_{-\infty}^{\infty}  \left[ (\xi-x) + a_2 (\xi-x)^2 + a_3 (\xi-x)^3 + a_4 (\xi-x)^4 \right] G(x, \xi, \tau) d \xi \nonumber \\
&= e^{-e^{x}} \left[ \frac{1}{2} \tau  e^{x} \left[2 + \tau - e^x \left(2 + \tau (7+e^x(e^x-6))\right)\right] - 1\right],  \nonumber \\
a_2 &= -\frac{1}{2} \left(e^x-1\right), \qquad a_3 = \frac{1}{6} \left(e^x \left(e^x-3\right)+1\right), \qquad a_4 = \frac{1}{24} \left(1-e^x \left(e^x \left(e^x-6\right)+7\right)\right). \nonumber
\end{align}
Fig.~\ref{figApp1} compares the "exact" values of $I_2(x)$, computed numerically, with those obtained from the analytical approximation in \eqref{appr1} for three values of $\tau = 0.02, 0.125, 0.3$. These values correspond to typical scenarios: a) $\sigma(t) = \sigma = 0.2, T = 1$; b) $\sigma(t) = \sigma = 0.5, T = 1$; and c) $\sigma(t) = \sigma = 1, T = 0.6$. Since the "exact" and approximated values are very close, in Fig.~\ref{figApp1} we display their difference. It can be seen that even in case c) the difference is less than one cent, therefore this approximation can be surely used in practice.

The second and third integrals in \eqref{modA} can be computed with linear complexity by using the Fast Gauss Transform (FGT), \cite{FGT2010}.  This efficiency arises from the Gaussian form of the Green's function (the kernel), while the remaining integrands are known in closed form (see above). Moreover, for the integral
\begin{equation} \label{PhiInt}
\int_0^\tau \int_{-\infty}^{\infty} \Theta(\nu,\xi) G(x, \xi, \tau-\nu) d\xi d\nu
\end{equation}
an approximation analogous to \eqref{appr1} can be constructed following the same approach:
\begin{enumerate}
\item Add and subtract a term proportional to $e^{-e^x}/\alpha(\tau)$.
\item Expand $\Theta(\nu,\xi)$ in a Taylor series around $\xi = x$ up to, say, fourth order.
\end{enumerate}
This procedure allows the integral in \eqref{PhiInt} to be evaluated in closed form.

One more important point for computing these integrals numerically is that, by the definition of $\alpha(t)$ in \eqref{tr1}, we have
\begin{equation} \label{alphaInt}
\alpha(t) = e^{\frac{1}{2} \int_T^t \sigma^2(k) - \int_T^t a(k) dk}.
\end{equation}
Based on the definition of $a(t)$ in \eqref{divDef}, the second integral in \eqref{alphaInt} can be rewritten as
\begin{equation}
\int_T^t a(k) dk = \int_T^t [r(k) - q(k) dk - \sum_{i=1}^{N_p} d_i \bm{1}_{t_{i} \ge t}.
\end{equation}

\subsection{Transition density function} \label{trDensity}

An alternative approach to pricing European options, when the underlying stock $S_t$ follows the dynamics in \eqref{GBM-41}, is to derive the transition density function $p(t,s | 0,S)$ of the process and then compute the option price by evaluating the first expectation in \eqref{decompGen}. Furthermore, if the EB $S_B(t)$ is known, the density function also enables the computation of the second expectation in \eqref{decompGen2}. In other words, given the EB, knowledge of the density resolves the problem of pricing American Put options with discrete dividends.

It is well-known that the density function $p(t,s | 0,S)$ of the stochastic process is the solution of the corresponding Fokker-Planck equation, see \cite{gardiner2004handbook,Kampen1992} among others. For our process in \eqref{GBM-41} it reads
\begin{equation} \label{FP}
\fp{}{t} p(t,s | 0,S) = - \fp{}{s} [ \mu(t,s) p(t,s | 0,S)] + \frac{1}{2} \sop{}{s}[\sigma^2(t) p(t,s | 0,S)],
\end{equation}
where $s$ denotes the current state of the stochastic process at time $t$, and $S$ is the initial state at time $t=0$. It should be solved subject to the initial condition
\begin{equation} \label{tcP}
p(0,s | 0,S) = \delta(s-S),
\end{equation}
and homogenous boundary conditions
\begin{equation} \label{bcP}
p(t,0 | 0,S) = p(t,s | 0,S)|_{s \uparrow \infty} = 0.
\end{equation}
Expanding \eqref{FP}, we obtain
\begin{align}
\fp{p}{t} &= \dfrac{1}{2}\sigma^2(t) s^2 \sop{p}{s} +  [\bar{a}(t) s - b(t)] \fp{p}{s} - \bar{r}(t) p, \\
\bar{r}(t) &= a(t) - \sigma^2(t), \qquad \bar{a}(t) = 2 \sigma^2(t) - a(t). \nonumber
\end{align}
By doing a change of variables similar to that in \eqref{tr1}
\begin{align} \label{tr1P}
s \to \frac{K}{\bar{\alpha}(t)}e^x, \quad p(t,x|0,x_0) &= W(t,x|0,x_0)e^{\int_0^t \bar{r}(k) dk}, \quad \bar{\alpha}(t) = e^{\int_0^t \left[ \frac{3}{2} \sigma^2(k) - a(k)\right] dk}, \\
(t, x) &\in [0,T]\times \mathbb{R}, \quad x_0 = \log(S/K). \nonumber
\end{align}
and then the transformation of the time similar to \eqref{tau}
\begin{equation} \label{tauP}
\tau = \frac{1}{2} \int_0^t \sigma^2(k) dk \ge 0,
\end{equation}
we arrive at \eqref{PDE2} with the source term $\Phi(\tau,x) = 0$. In the new variables, the boundary condition \eqref{bcP} and initial condition from \eqref{tcP} become
\begin{equation} \label{bc1P}
W(\tau,x|0.x_0)\big|_{x \to \infty} = W(\tau,x|0.x_0)\big|_{x \to -\infty} = 0, \quad W(0,x|0,x_0) = \delta(x-x_0).
\end{equation}

The \eqref{PDE2} has been already solved by using the method outlined in \cref{Europ}. Furthermore, because the initial condition in \eqref{tcP} holds and $\Phi(\tau,x)$ vanishes, the expressions in \cref{Europ} simplify significantly. To enhance transparency, a detailed application of \cref{proc} specifically for computing the transition density, is presented in Appendix~\ref{app1}.

\begin{remark}
Our analysis has thus far focused on Put options. However, the decomposition formula for American call options shares the same general form as \eqref{decompGen2} since follows from the change of variables formula of \cite{Peskir2005}. The key differences lie in the payoff function at maturity, and the structure of the exercise region, represented by the indicator function $\mathbf{1}_{S_u \in \mathcal{E}}$ in \eqref{decompGen2}. For Put options, the exercise region is $S_u \in [0, S_B(t)]$, whereas for Call options it becomes $S_u \in [S_B(t), \infty)$. Since the transition density remains identical for both Put and Call options, the pricing methodology for Call options follows the same approach as for Puts.
\end{remark}

\section{Computing the exercise boundary $S_B(t)$} \label{secEBPut}

When pricing American options using the decomposition formula, the standard method for determining the EB proceeds as follows. At the EB, the option value is known: for a Put, it is $P(t, S_B(t)) = K - S_B(t)$, and for a Call, it is $CP(t, S_B(t)) = S_B(t) - K$. By substituting $S = S_B(t)$ into \eqref{decompGen2}, we obtain a nonlinear equation for $S_B(t)$. However, solving this equation can be computationally intensive because the EEP involves a double integral: one over time and another arising from the expectation.

As pointed out in \cite{ItkinKitapbayev2025}, if one needs to numerically compute the integral in \eqref{decompGen2} (the EEP) for the Put, a more efficient approach is to treat each integrand as the price of an Up-and-Out barrier Call option with the upper barrier: $S_B(u)$, maturity $u$, the payoff $H(u, S(u))$, and a zero strike. Thus, given $S_B(t)$ computation of the EEP requires pricing two barrier options for each maturity $u \in [t,T]$.

Therefore, it could more efficient to use the method proposed in \cite{ItkinMuravey2024jd}. Below we shortly describe this method as applied to our problem of pricing American Put option.

\subsection{The exercise boundary for the American Put option} \label{ebPut}

Similar to \cref{Europ}, by a standard argument, \citep{ContVolchkova2005, klebaner2005} for the model in \cref{GBM-41,divDef}, the American Put option price $P(t,S)$ in the {\it continuation} region $S \in \calC = [S_B(t), \infty)$ solves a parabolic PDE \eqref{PDE}, subject to the boundary conditions
\begin{equation} \label{bcEB}
P(t,S_B(t)) = K - S_B(t), \qquad P(t,S)|_{S \uparrow \infty} = 0,
\end{equation}
and the terminal condition
\begin{equation} \label{tcEB}
P(T,S) = (K - S)^+ = 0,
\end{equation}
since $S_B(T) = K$.

By a series of transformations similar to that in \cref{tr1,tau}
\begin{align} \label{trEB}
S &\to \frac{K}{\alpha(t)}e^x, \quad P(t,x) = u(\tau,x)e^{\int_T^t r(k) dk}, \quad \alpha(t) = \exp \left[ \frac{1}{2} \int_{T}^t [\sigma^2(k) - 2 a(k)] dk \right],  \nonumber \\
\tau &= - \frac{1}{2} \int_T^{t} \sigma^2(k) dk \ge 0, \quad y(\tau) = \log\left( \alpha(\tau) \frac{S_B(\tau)}{K} \right),
\quad (\tau, x) \in [0,\tau(0)]\times [y(\tau),\infty).
\end{align}
the PDE in \eqref{PDE} can be transformed to \eqref{PDE2} with no source term and the moving left boundary $y(\tau)$
\begin{gather} \label{PDE_EB0}
\fp{u}{\tau} = \sop{u}{x}  - e^{-x} \gamma(\tau) \fp{u}{x}, \\
\begin{align*}
 u(0,x) &= u(\tau,x)\Big|_{x \uparrow \infty} = 0, \quad u(\tau,y(\tau)) = K \beta(\tau) \left( \alpha(\tau) - e^{y(\tau)} \right) \nonumber \\
\beta(\tau) &= e^{\tau + 2 \int_0^\tau \left( \frac{r(k)}{\sigma^2(\tau)}\right) dk - \rho(\tau)}, \quad \alpha(\tau) = e^{-\tau + \rho(\tau) }, \quad \rho(\tau) = \int_0^\tau \frac{2 a(k)}{\sigma^2(k)}.
\end{align*}
\end{gather}
Further, we make another change of the dependent variable
\begin{align} \label{chgU}
U(\tau,x) &= u(\tau,x) - z(\tau,x) e^{- \tau (y(\tau)-x)^2}, \qquad z(\tau,x) = K \beta(\tau) \left( \alpha(\tau) - e^{x} \right),
\end{align}
where the reason for the specific form of the subtracted term in the right-hands part of \eqref{chgU} will become clear in \cref{singSect}.

The transformation in \eqref{chgU} reduces the problem in \eqref{PDE_EB0} to that with homogeneous boundary conditions
\begin{gather} \label{PDE_EB}
\fp{U}{\tau} = \sop{U}{x}  - e^{-x} \gamma(\tau) \fp{U}{x} + h(\tau,x), \qquad
U(0,x) = - z(0,x), \qquad U(\tau,x)\Big|_{x \uparrow \infty} = U(\tau,y(\tau)) = 0, \\
\begin{align*}
h(\tau,x) &=  e^{- \tau (x-y(\tau))^2} \left\{ \gamma(\tau) \zeta(\tau,x) + \eta(\tau,x) \right\}, \qquad
\zeta(\tau,x) = 2 \tau z(\tau,x) (x - y(\tau))e^{-x} + K \beta(\tau), \nonumber \\
\eta(\tau,x) &= z(\tau,x) \left[ (x-y(\tau))^2 (1 + 4 \tau^2)  - 2 \tau y'(\tau) (x-y(\tau)) - (2 \tau + \bar{r}(\tau)) \right]
\nonumber \\
&+ K \beta(\tau) e^x \left[4 \tau (x-y(\tau)) -  \rho'(\tau)\right], \qquad \bar{r}(\tau) = \frac{2 r(\tau)}{\sigma^2(\tau)}. \nonumber
\end{align*}
\end{gather}

This problem was already solved in \cite{ItkinMuravey2024jd}, Section~1.2.2. A comparison of the frameworks shows that our case corresponds to setting their function $g(\tau) = 0$, while their source term $\lambda(\tau, x)$ becomes
\begin{equation} \label{lambda}
\lambda(\tau, x) = - e^{-x} \gamma(\tau) \fp{U}{x} + h(\tau,x).
\end{equation}
Although our source term depends on $\fp{U}{x}$, this does not pose any issues due to the generalized Duhamel’s principle, see \cite{Itkin2024jd, Hunter2014}.

Accordingly, based on \cite{ItkinMuravey2024jd}, the solution of \eqref{PDE_EB} reads
\begin{align} \label{solPut}
U(\tau, x) &=  \frac{1}{2\sqrt{\pi \tau}} \int_{y(0)}^{\infty} U(0,\xi) \left[e^{-\frac{( \xi - x)^2}{4\tau}}
-  e^{-\frac{(\xi + x -2 y(\tau))^2}{4\tau}} \right] d\xi \\
&- \int_{0}^{\tau} \frac{\Psi(s, y(s))}{2\sqrt{\pi(\tau - s)}} \Bigg[ e^{-\frac{(x - y(s))^2}{4(\tau - s)}} - e^{-\frac{(x - 2y(\tau) + y(s))^2}{4(\tau - s)}} \Bigg]ds \nonumber \\
&+ \int_{0}^{\tau} \int_{y(s)}^{\infty} \frac{\lambda(s, \xi)}{2\sqrt{\pi(\tau - s)}} \left[ e^{-\frac{(\xi - x)^2}{4(\tau - s)}} - e^{-\frac{(\xi - 2y(\tau) + x)^2}{4(\tau - s)}} \right] d\xi ds, \nonumber
\end{align}
where $\Psi(\tau, y(\tau))$ is the gradient of the solution at the moving boundary $y(\tau)$
\begin{equation}
\Psi(\tau, y(\tau)) =  \fp{U(\tau, x)}{x} \Big |_{x = y(\tau)}.
\end{equation}
Due to the smooth-pasting condition for American options, see \cite{kwok2008mathematical} among others, $P_S(t, S_B(t)) = -1$. Accordingly, after doing some algebra, we obtain
\begin{align} \label{sub}
\Psi(\tau, y(\tau)) &= 0.
\end{align}
However, strictly speaking, the smooth pasting condition is not required here since our problem is confined to the continuation region. In other words, \eqref{sub} needs only hold within the continuation region, without necessarily extending to the exercise region. Consequently, the EB may not be continuous (and not monotonic) in $S$ and could exhibit a jump at the ex-dividend date, see \cite{vellekoop2011integral}.

The first integral in \eqref{solPut} can be taken in closed form to yield
\begin{align}
I_0 &= \frac{1}{2\sqrt{\pi \tau}} \int_{y(0)}^{\infty} U(0,\xi) \left[e^{-\frac{( \xi - x)^2}{4\tau}} -  e^{-\frac{(\xi + x -2 y(\tau))^2}{4\tau}} \right] d\xi = - \frac{1}{2}K \Bigg[ \erf\left(\frac{x}{2 \sqrt{\tau }} \right) + \erf\left(\frac{x-2 y(\tau )}{2 \sqrt{\tau }} \right) \\
&+ e^{\tau + x} \left(\erfc\left(\frac{2 \tau +x}{2 \sqrt{\tau }}\right)-2\right)  +
e^{x + \tau +2 y(\tau )} \erfc\left(\frac{-2 \tau +x-2 y(\tau )}{2 \sqrt{\tau }} \right) \Bigg]. \nonumber
\end{align}

Since in our case $\lambda(s,\xi) \propto U'_\xi(s,\xi)$, \eqref{solPut} is actually not the solution but rather an integral Volterra equation of the first kind for $U(s,\xi)$. Indeed, with allowance for \eqref{lambda}, after integrating by parts the last integral in \eqref{solPut} reads
\begin{gather} \label{simplU}
\calJ(\tau,x) = \calJ_1(\tau,x) + \calJ_2(\tau,x), \qquad \calJ_2(\tau,x) = \int_{0}^{\tau} \gamma(s) (J_1 + J_2) ds, \\
\begin{align*}
\calJ_1(\tau,x) &= \int_0^\tau \int_{y(s)}^\infty e^{- s (\xi-y(s))^2} \frac{\eta(s,\xi)}{2\sqrt{\pi(\tau - s)}} \left[ e^{-\frac{(\xi - x)^2}{4(\tau - s)}} - e^{-\frac{(\xi - 2y(\tau) + x)^2}{4(\tau - s)}} \right] d\xi ds, \\
J_2 &=  \int_{y(s)}^\infty e^{- s (\xi-y(s))^2} \frac{\zeta(s,\xi) }{2\sqrt{\pi(\tau - s)}} \left[ e^{-\frac{(\xi - x)^2}{4(\tau - s)}} - e^{-\frac{(\xi - 2y(\tau) + x)^2}{4(\tau - s)}} \right] d\xi, \\
J_1 &= \int_{y(s)}^{\infty} e^{-\xi} U(s,\xi) \Bigg\{- \frac{1}{2\sqrt{\pi(\tau - s)}}
\left[ e^{-\frac{(x-\xi )^2}{4 (\tau-s )}} - e^{-\frac{(\xi + x-2 y(\tau ) )^2}{4 (\tau-s )}} \right] \\
&+ \frac{1}{4\sqrt{\pi(\tau - s)^3}} \left[(x-\xi)e^{-\frac{(x-\xi )^2}{4 (\tau-s )}} + (x + \xi - 2y(\tau)) e^{-\frac{(\xi +x-2 y(\tau ))^2}{4 (\tau-s )}}  \right] \Bigg\} d\xi.
\end{align*}
\end{gather}
Since \eqref{solPut} contains two unknown functions, $U(\tau, x)$ and $y(\tau)$, a second equation is needed to determine the early exercise boundary (EB). Following \cite{ItkinMuravey2024jd}, we derive this by differentiating both sides of \eqref{solPut}. Differentiation with respect to $x$ yields the option's Delta, while differentiation with respect to $\tau$ yields the options's theta\, \footnote{For the underlying GBM process with constant coefficients, an integral equation for $y(\tau)$ using the option's Theta was derived in \cite{GoodmanOstrov2002, ZhuHeLu2018} using a different approach. However, their approach cannot be used for models with time-dependent coefficients.}. We proceed by differentiating with respect to $x$ and then evaluating at $x \to y(\tau)$, which gives (see Appendix~\ref{appPsi}):
\begin{align} \label{VolEB1}
0 &= \frac{1}{\sqrt{\pi \tau}}\int_{y(0)}^\infty U'_\xi(0,\xi) e^{-\frac{(\xi - y(\tau))^2}{4\tau}} d\xi +  \int_0^\tau \int_{y(s)}^\infty \frac{e^{-\frac{(\xi - y(\tau))^2}{4(\tau - s)}}}{2\sqrt{\pi(\tau - s)}} \frac{\xi - y(\tau)}{\tau-s}  \eta(s,\xi) e^{-s(\xi-y(s))^2} d\xi ds, \\
&- \int_0^\tau \gamma(s) \int_{y(s)}^{\infty} \left[ e^{-\xi} U'_\xi(s,\xi) - \zeta(s,\xi) e^{-s (\xi-y(s))^2} \right] \frac{\xi - y(\tau)}{2\sqrt{\pi (\tau - s)^3}} e^{-\frac{(\xi - y(\tau))^2}{4(\tau-s)}} d\xi ds, \nonumber
\end{align}

The first integral in \eqref{solPut} can be taken in closed form, and since $y(0) = 0$ this yields
\begin{align}
J_0 &= -\int_0^\infty \frac{ e^{-\frac{(\xi -y(\tau))^2}{4\tau}}}{\sqrt{\pi\tau}} \fp{f(0,\xi)}{\xi} \, d\xi = K e^{\tau + y(\tau)} \left[ 1 + \erf\left( \frac{2\tau + y(\tau)}{2 \sqrt{\tau}}\right) \right].
\end{align}

The impact of dividends is incorporated through two key functions: $\gamma(\tau)$, which accounts for discrete cash dividends, and $\alpha(\tau)$, which models discrete proportional dividends. Since by definitions in \eqref{PDE2}, \eqref{divDef}
\begin{equation} \label{gamma2}
\gamma(s) = \frac{2\alpha(s)}{K \sigma^2(s)}  \sum_{j=1}^{N_d} D_j \delta(s - \tau_{j}) \bm{1}_{s \le \tau_{j}},
\end{equation}
\eqref{VolEB1} reduces to the equation
\begin{align} \label{VolEB2}
- K e^{\tau + y(\tau)} & \left[ 1 + \erf\left( \frac{2\tau + y(\tau)}{2 \sqrt{\tau}}\right) \right] = \int_0^\tau \int_{y(s)}^\infty \frac{e^{-\frac{(\xi - y(\tau))^2}{4(\tau - s)}}}{2\sqrt{\pi(\tau - s)}} \frac{\xi - y(\tau)}{\tau-s}  \eta(s,\xi) e^{-s (\xi-y(s))^2} d\xi ds \\
&- \sum_{j=1}^{N_d} \bm{1}_{\tau_{j} < \tau} \frac{2 \alpha(\tau_{j})}{\sigma^2(\tau_{j})} D_j \Bigg\{
- \int_{y(\tau_{j})}^{\infty} \zeta(\tau_{j}, \xi) e^{- \tau_{j}(\xi-y(\tau_{j}))^2}
\frac{\xi - y(\tau)}{2\sqrt{\pi (\tau - \tau_{j})^3}} e^{-\frac{(\xi - y(\tau))^2}{4(\tau-\tau_{j})}} d\xi \nonumber \\
&+ \int_{y(\tau_{j})}^{\infty} e^{-\xi} U'_\xi(\tau_{j},\xi) \frac{\xi - y(\tau)}{2\sqrt{\pi(\tau-\tau_{j})^3}} e^{-\frac{(\xi-y(\tau))^2}{4(\tau-\tau_{j})}} d\xi \Bigg\}. \nonumber
\end{align}
The last integral in this equation can be further simplified, as in \eqref{simplU}.

In case $\tau = \tau_{j}$, in \eqref{VolEB2} the last integral under the sum vanishes because
\begin{align}
I &= \lim_{\tau \to \tau_{j}} \frac{\xi - y(\tau)}{2\sqrt{\pi(\tau-\tau_{j})^3}} e^{-\frac{(\xi-y(\tau))^2}{4(\tau-\tau_{j})}} = \lim_{\tau \to \tau_{j}} \frac{1}{2\sqrt{\pi(\tau - s)}} \partial_x \left[ e^{-\frac{(\xi - x)^2}{4(\tau - s)}} - e^{-\frac{(\xi - 2y(\tau) + x)^2}{4(\tau - s)}} \right]_{\substack{x = y(\tau) \\ s = \tau_{j}}} \\
&= \partial_x \left[ \delta(\xi-x) - \delta(\xi + x - 2 y(\tau)) \right] \Big|_{x = y(\tau)}
= -2 \delta'(\xi - y(\tau)), \nonumber
\end{align}
and
\begin{align}
-2 \int_{y(\tau)}^{\infty} e^{-\xi} U'_\xi(\tau,\xi) \delta'(\xi - y(\tau)) d\xi = \partial_\xi \left[e^{-\xi} U'_\xi(\tau,\xi) \right]_{\xi = y(\tau)} = 0.
\end{align}
Accordingly, we use $\bm{1}_{\tau_{j} < \tau}$ instead of $\bm{1}_{\tau_{j} \le \tau}$ when doing summation.

Thus, at given $\tau_j{-}$, and $\tau \le \tau_{j}$ \eqref{VolEB2} reduces to a nonlinear Volterra equation of the first kind for $y(\tau)$, as all terms in the summation of \eqref{VolEB2} are already known at time $\tau$. However, once $\tau$ exceeds $\tau_{j}$, the sum in the right-hands part of \eqref{VolEB2} will further contain a extra term dependent on $U(\tau_{j},\xi)$ which is not known yet. Fortunately, it can be found from \eqref{solPut}. The latter equation after substituting $\gamma(s)$ from \eqref{gamma2} and $\tau = \tau_{j}$ becomes
\begin{align} \label{solPut2}
U(\tau_{j}, x) &= \calJ_1(\tau_{j},x) + \sum_{i=1}^{N_d} \bm{1}_{\tau_{i} < \tau_{j}} \frac{2 \alpha(\tau_{i})}{\sigma^2(\tau_{i})} D_i \calJ_3(\tau_{j}, \tau_{i},x) + \frac{2 \alpha(\tau_{j})}{\sigma^2(\tau_{j})} D_j \calJ_3(\tau_{j}, \tau_{j},x), \\
\calJ_3(\tau_{j}, \tau_{i},x) &= \int_{y(\tau_{i})}^{\infty} e^{-\xi} U(\tau_{i},\xi) \left\{ - \frac{1}{2\sqrt{\pi(\tau_{j} - \tau_{i})}} \left[ e^{-\frac{(x-\xi )^2}{4 (\tau_{j}-\tau_{i})}} - e^{-\frac{(\xi +x-2 y(\tau_{j} ) )^2}{4 (\tau_{j}-\tau_{i})}} \right] \right. \nonumber \\
&+ \left. \frac{1}{4\sqrt{\pi(\tau_{j} - \tau_{i})^3}} \left[(x-\xi)e^{-\frac{(x-\xi )^2}{4 (\tau_{j}-\tau_{i})}} + (x + \xi - 2y(\tau_{j})) e^{-\frac{(\xi + x-2 y(\tau_{j} ))^2}{4 (\tau_{j}-\tau_{i})}}  \right] \right\} d\xi \nonumber \\
&+ \int_{y(\tau_{i})}^{\infty} \zeta(\tau_{i},\xi) e^{- \tau_{i} (\xi-y(\tau_{i}))^2} \frac{\xi - y(\tau_{j})}{2\sqrt{\pi (\tau_{j} - \tau_{i})^3}} e^{-\frac{(\xi - y(\tau_{j}))^2}{4(\tau_{j}-\tau_{i})}} d\xi, \qquad i < j, \nonumber \\
\calJ_3(\tau_{j}, \tau_{j},x) &= - \zeta_\xi(\tau_{j}, y(\tau_{j})) - e^{-x} U'_x(\tau_{j},x). \nonumber
\end{align}
The expression for $\calJ_3(\tau_{j}, \tau_{j},x)$ arises from the definition of $\calJ_3(\tau_{j}, \tau_{i},x)$ having in mind that  at $i=j$ the expression in first square brackets is a difference of two delta functions, the second square brackets gives rise to a difference of the first derivatives of delta functions, the third integral contains a derivative of the delta function as an integrand, and also $x \ge y(\tau)+0$. In more detail, see Appendix~\ref{appTj}.

Thus, $U(\tau_{j},x)$ solves an ordinary differential equation (ODE)
\begin{gather} \label{Phi}
U(\tau_{j},x) + w_j e^{-x} U'_x(\tau_{j},x) = \Phi(\tau_{j},x), \\
\begin{align*}
\Phi(\tau_{j},x) &= \calJ_1(\tau_{j},x) - w_j \zeta_\xi(\tau_{j}, y(\tau_{j})) \\
&+ \sum_{i=1}^{N_d} \bm{1}_{\tau_{i} < \tau_{j}} \frac{2 \alpha(\tau_{i})}{K \sigma^2(\tau_{i})} D_i \calJ_3(\tau_{j}, \tau_{i},x), \qquad w_j = \frac{2 \alpha(\tau_{j})}{K \sigma^2(\tau_{j})} D_j,
\end{align*}
\end{gather}
subject to the zero boundary condition at $x = y(\tau_{j})$. The solution reads
\begin{align} \label{Usol}
U(\tau_{j},x) &= \frac{1}{w_j} \int_{y(\tau_{j})}^x \Phi(\tau_{j},k) \exp\left( k + \frac{e^k - e^x}{w_j}\right) dk,
\end{align}
and hence
\begin{align} \label{Uprime}
U'_x(\tau_{j},x) &= \frac{e^x}{w_j} \left[ \Phi(\tau_{j},x) - \frac{1}{w_j}
e^{-e^{x}/w_j} \int_{y(\tau_{j})}^x e^{k + e^k/w_j} \Phi(\tau_{j},k)dk \right].
\end{align}

Since \eqref{solPut2} provides an explicit expression for $U(\tau_{j}, x)$, it can further be substituted into \eqref{VolEB2}, so \eqref{VolEB2} again becomes a nonlinear Volterra equation of the first kind for $y(\tau)$. The inclusion of an additional term in \eqref{VolEB2} when $\tau$ reaches $\tau_{j}$ causes the EB $y(\tau)$ to exhibit a jump at $\tau_j$. In the zero-order approximation, the magnitude of this jump is proportional to $D_j$, i.e $S_B(t_{j})  = S_B(t_{j}) e^{D_j \eta(\tau_j,y(\tau_j))}$, where $\eta(\tau_j,y(\tau_j))$ is a function that can be explicitly determined using \eqref{VolEB2} and \eqref{solPut2}.

Because $y(\tau)$ is computed sequentially backward in time, the integrals over $s$ contain only one unknown term corresponding to the most recent time interval $\Delta \tau = (\tau_{i1}, \tau_i]$. After all integrands are evaluated at a single point $\tau_i$ (they are already known at previous points $\tau_k, \, k < i$, \eqref{VolEB1} can be solved numerically to obtain $y(\tau_i)$. The algorithm repeats until convergence is achieved within the desired tolerance. As reported in \cite{Itkin2024jd}, where a similar system of equations was solved using an analogous method, the latter typically converges within 5–6 iterations. Since all integrals in \eqref{solPut2} involve Gaussian kernels, the FGT can be efficiently applied to compute them.

It is important to emphasize that $U(\tau,x)$ is not the price of the American option given by \eqref{decompGen2}. Rather, it represents the American put option price within the continuation region $\calC$, where it coincides with the price of its European counterpart, subject to the boundary and terminal conditions in \cref{bcEB,tcEB}. Using the approach outlined in \cref{Europ}, the European put price can be obtained for the entire domain $S \in [0, \infty)$ as well as the transition density (see \cref{trDensity}). Once the EB, the European option price, and the transition density are determined, they can be substituted into \eqref{decompGen} to compute the full American put option price.

\subsubsection{Singularity in $y'(\tau)$ at $\tau \to 0$} \label{singSect}

As noted in \cite{ZhuHeLu2018}, almost all the existing integral equations for optimal exercise boundary of the American option known in the literature are exposed to a singularity at $\tau=0$ due to the presence of the negative indefinite slope of the EB $y'(0) = - \infty$. This is important because numerical methods (like finite difference or quadrature methods) approximate solutions by breaking time into small discrete steps. A singularity represents an infinite value within the very first, most critical time step. This corrupts the numerical solution from the outset.

The error introduced at $\tau=0$ due to various numerical approximations does not remain isolated. It propagates and amplifies through the entire numerical scheme as the algorithm steps backward in time (i.e., as $\tau$ increases). This leads to a computed EB that can be significantly inaccurate for times well before expiry. Also, a key requirement for a numerical method is that as the time grid is made finer (smaller time steps), the solution should converge to the true answer. The presence of a singularity often prevents this convergence, or makes it impractically slow. The solution may oscillate wildly or produce nonsensical results as the step size decreases.

At the first glance, our integral equation for the exercise boundary (EB) in \eqref{VolEB2} appears to suffer from the same problem. It contains the term $y'(s)$ which is defined by \eqref{PDE_EB} and is integrated from a lower limit of $s=0$, suggesting a potential singularity. However, we aim to demonstrate that by a change of variables this singularity can be removed.

It is well-known, see e.g., \cite{Evans2002}, that for a GBM process with constant coefficients a short-time asymptotic of $S_B(\nu)$ at $\nu \to 0^+$ where $\nu = \sigma^2 (T-t)$, takes the following forms
\begin{align} \label{asympt}
SB(\nu) &\sim
\begin{cases}
K - K \sqrt{\nu \ln \left(\frac{\sigma^4}{8 \pi \nu (r-q)^2}\right)}, & 0 \le q < r, \\
K - K \sqrt{2 \nu \ln \left(\frac{\sigma^2}{4 \sqrt{\pi} q \nu} \right)}, & q = r, \\
\frac{r}{q} K (1-\bar{\alpha} \sqrt{2 \nu}), & q > r.
\end{cases}
\end{align}
Here, $\bar{\alpha}$ is a numerical constant which satisfies a known transcendental equation. Accordingly, introducing $\bar{\nu} = \sigma^2 \nu$, we obtain
\begin{align}
SB'(\nu) &\sim K
\begin{cases}
\sqrt{-\log(\bar{\nu})/\bar{\nu}}, & 0 \le q \le r, \\
- \frac{\bar{\alpha}}{2\sqrt{\bar{\nu}}}, & q > r,
\end{cases}
\end{align}
so $S'_B(\bar{\nu})$ is singular when $\bar{\nu} \to 0$. Accordingly, by the definition of $y(\tau)$ in \eqref{trEB} we have
\begin{equation} \label{yPrime}
y'(\tau) = -1 + \rho'(\tau) + \frac{S'_B(\tau)}{S_B(\tau)}, \qquad S'_B(\tau) = S'_B(\nu)\fp{\nu}{\tau} = 2 S'_B(\nu) + O(\nu),
\end{equation}
and, thus, $y'(\tau)$ is also singular at $\tau \to 0$.

The definitions of $y(\tau)$ in \eqref{trEB} and the asymptotic behavior of $S_B(\tau)$ in \eqref{asympt} imply
\begin{align} \label{asympSim}
y(\tau) &\sim -\sqrt{-\tau \log(\tau)}, \qquad y'(\tau) \sim - \sqrt{-\log(\tau)/\tau}.
\end{align}

A careful analysis shows that the term $y'(s)$ appears only within the function $\eta(s,\xi)$ in the double integral in \eqref{VolEB2}, which can be represented as
\begin{align} \label{defG}
I &= \int_0^\tau G(\xi, s, \tau)  ds, \qquad
G(\xi, s, \tau) = \int_{y(s)}^\infty \frac{e^{-\frac{(\xi - y(\tau))^2}{4(\tau - s)}}}{2\sqrt{\pi(\tau - s)}} \frac{\xi - y(\tau)}{\tau-s}  \eta(s,\xi) e^{- s (\xi-y(s))^2} d\xi.
\end{align}
But $y'(s)$ enters the definition of $\eta(s,\xi)$ in \eqref{PDE_EB} only as a product $s y'(s)$. Based on \eqref{asympSim}, one can see that
\begin{equation}
\lim_{s \to 0} s y'(s) \sim \lim_{s \to 0} y(s) = 0.
\end{equation}
Therefore, our integral equations have no singularity at $s \to 0$. This is achieved by a special choice of the correcting function $z(\tau,x) e^{- \tau (y(\tau)-x)^2}$ in \eqref{chgU}.

It is important to note that $G(\xi,x,\tau)$ also has no singularity at $s = \tau$ because in this case \begin{align}
\lim_{s \to \tau} \frac{e^{-\frac{(\xi - y(\tau))^2}{4(\tau - s)}}}{2\sqrt{\pi(\tau - s)}} \frac{\xi - y(\tau)}{\tau-s}   = - 2 \delta'(\xi - y(\tau)).
\end{align}

This results demonstrates that our method is free of singularities in the temporal variable $\tau$. This finding aligns with the singularity-free results of \cite{GoodmanOstrov2002,ZhuHeLu2018}, who addressed the cases of no dividends and time-inhomogeneous GBM model by constructing an integral equation for $y(\tau)$ via the option's theta. A critical distinction, however, is that their methodology is incompatible with models featuring time-dependent coefficients.

Finally, note that solving \eqref{VolEB2} requires knowledge of $y'(s)$ for all $s \in [0, \tau]$. While we have determined its value at $s=0$, at other points $s \in (0, \tau]$ it can be approximated by using a finite difference scheme, which is naturally compatible with the temporal grid used to solve the integral equation numerically.

\subsubsection{The no-dividends case}

In case of no dividends (i.e., $\gamma(t) = q(t) = 0$), \eqref{VolEB2} transforms to
\begin{align} \label{VolEB-a1}
- K e^{\tau + y(\tau)} & \left[ 1 + \erf\left( \frac{2\tau + y(\tau)}{2 \sqrt{\tau}}\right) \right] = \int_0^\tau \int_{y(s)}^\infty \frac{e^{-\frac{(\xi - y(\tau))^2}{4(\tau - s)}}}{2\sqrt{\pi(\tau - s)}} \frac{\xi - y(\tau)}{\tau-s}  \eta(s,\xi) e^{-s (\xi-y(s))^2} d\xi ds.
\end{align}
It it is easy to see that the left-hands part of \eqref{VolEB-a1} is non-positive.

By the definitions in \eqref{trEB} we have $y(\tau) \le 0$ since
\begin{align} \label{a2}
\alpha(\tau) &= e^{-\tau + \rho(\tau)}, \qquad \rho(\tau) = \int_0^\tau \frac{2 a(k)}{\sigma^2(k)}, \qquad
y(\tau) = -\tau + \rho(\tau) + \log(S_B(\tau)/K),
\end{align}
and for the American Put option $K \ge S_B(\tau)$.
On the other hand, for the American Put option $S_B'(\tau) < 0$, and hence
\begin{equation}
y(s) - y(\tau) = (\tau -s) \left[1 - \frac{1}{\tau-s} (\bar{r}(\tau) - \bar{r}(s)) \right] + \log \left( \frac{S_B(s)}{S_B(\tau)} \right) > 0,
\end{equation}
if $\tau > s$ and $\bar{r}'(\tau) < 1$. For instance, if $r(t), \sigma(s)$ are constants, this means
$2r/\sigma^2 < 1$. Moreover, the inequality $y(s) - y(\tau) > 0$ still holds if $\fp{\bar{r}(s)}{s} < 1 - \fp{\log(S_B(s))}{s} > 1$ for all $0 \le s \le \tau$. Thus, the sign of the integral in the right-hands part of \eqref{a2} coincides with that of $\eta(s,\xi)$.

This statement holds because $\xi - y(\tau) > 0$ even when $\xi = y(s)$. Furthermore, the integrand decays rapidly for large $\xi$ due to the product of two Gaussian exponential terms. Consequently, the integral's dominant contribution comes from the region near the lower limit, $\xi = y(s)$. At this limit we have
\begin{align} \label{pAtY}
\eta(s, y(s)) &= - (2 s + \br'(s)) z(s, y(s)) - K \beta(s) e^{y(s)} \rho'(s) = - 2 s z(s, y(s)) - K e^{\bar{r}(s)} \bar{r}'(s), \\
z(s, y(s)) &= K e^{\bar{r}(s)} \left(1 - \frac{S_B(s)}{K} \right) \ge 0, \qquad \br(\tau) = \int_0^\tau \frac{2 r(k)}{\sigma^2(k)}. \nonumber
\end{align}
This result holds because for no dividends $\bar{r}(s) = \rho(s)$.

Thus, $\eta(s,y(s)) \leq 0$ and both sides of \eqref{VolEB-a1} are negative. The existence of a real-valued solution $y(\tau) < 0$ therefore follows, a conclusion supported by numerical results.

However, when $r < 0$, the derivative $S'_B(\tau)$ becomes positive. This occurs because the holder of the American Put has a stronger incentive to exercise longer-dated options early to avoid the cost of holding cash with a negative yield. In other words:
\begin{itemize}
\item When $r > 0$, the opportunity cost of forfeiting interest on $K$ discourages early exercise for longer-dated options, resulting in a lower boundary for higher $\tau$.

\item In contrast, when $r < 0$, the "opportunity benefit" of receiving $K$ early to avoid its negative yield encourages early exercise for longer-dated options, resulting in a higher boundary for higher $\tau$.
\end{itemize}

Because $S'_B(s) > 0$, it follows that $y'(\tau) > 0$ and hence, $y(s) - y(\tau) < 0$. while $\eta(s,y(s)$ remains same.
Consequently, the integrand now becomes non-negative for all $\xi \in [y(s), \infty)$. This, in turn, leads to a contradiction: the left-hand side of \eqref{VolEB-a1} is non-positive, while its right-hand side is non-negative. Thus, the integral equation possesses no real-valued solutions except of $y(\tau) = -\infty$, or $S_B(\tau) = 0$.

This means that the EB for an American Put doesn't exist (or, is undefined) for $r < 0$ and the whole domain $S \in [0, \infty)$ is just a continuation region. Although the absence of the EB in this case is well-established, it is typically argued from a financial perspective, while here we derive it mathematically through an analysis of \eqref{VolEB-a1}.

The absence of an EB implies that early exercise is never optimal for the American Put option. Consequently, its price equals that of the corresponding European option, and the EEP is zero.

At the end, notice that the double integral in the right-hands part of \eqref{VolEB-a1} can be transformed to a single integral as the internal integral in $\xi$ can be taken in closed form, see Appendix~\ref{appGA}.

\subsection{Volterra equations with weak singularities} \label{weakSing}

It turns out that \eqref{solPut} is a Volterra integral equation with weakly singular kernels. Indeed, denote $\calD$ to be the domain of definition of the variable $\tau$. In our case $\calD = [0,\tau(0)]$. The equation of the form
\begin{equation} 	\label{VIE_weak_kernel-32}
u(\tau) = f(\tau) + \int_0^\tau k_\alpha(t-s) K(\tau,s) u(s) ds.
\end{equation}
\noindent with $(\tau,s) \in D, \ K(\tau,s) \in \calC(D)$, and the weakly singular function $k_\alpha(\tau-s)$
 \begin{equation} \label{k_alpha_def-32}
k_\alpha(\tau-s) =
\begin{cases}
(\tau-s)^{\alpha - 1}, & 0 < \alpha < 1, \\
\log(\tau-s), & \alpha = 0,
\end{cases}
\end{equation}
\noindent is called the Volterra integral equation\index{Volterra integral equation} with a weakly singular kernel.

It is easy to see that, for instance, \eqref{VolEB1} obeys the definition in \eqref{VIE_weak_kernel-32} if in the double integral we take an integrand at the lower limit $\xi=y(s)$, so
\begin{align} \label{defWSI1}
k_{1/2}(\tau-s) &= \frac{1}{\sqrt{\tau-s}}, \qquad K(\tau,s) = \frac{\eta(s, y(s))}{\sqrt{\pi}} \frac{y(\tau)- y(s)}{\tau -s} e^{-\frac{(y(\tau) - y(s))^2}{4(\tau-s)}}.
\end{align}

To illustrate regularity of $K(\tau,s)$ in \eqref{defWSI1}, observe that $K(\tau,s)$ is regular if $s \neq \tau$. Since $y(\tau) \in \calC^1$, the limit $s \to \tau$ can be computed as follows
\begin{equation*}
\lim_{s \to \tau} \frac{\eta(s, y(s))}{\sqrt{\pi}} \frac{y(\tau)- y(s)}{\tau -s} e^{-\frac{(y(\tau) - y(s))^2}{4(\tau-s)}} = \eta(\tau, y(\tau)) \frac{y'(\tau)}{\sqrt{\pi}} < \infty,
\end{equation*}
etc. At the same time the existence and uniqueness of the solution of \eqref{VIE_weak_kernel-32} still holds. Since $K(\tau,s) \in \calC^0, f \in \calC^0$ and $0 < \alpha < 1$,  \eqref{VIE_weak_kernel-32} possesses a unique solution $u \in \calC^0$ given by
\begin{equation}
u(\tau) = f(\tau) + \int_0^\tau R_\alpha(\tau,s) f(s) ds.
\end{equation}
Here $R_\alpha(\tau,s)$ is the resolvent kernel corresponding to $K_\alpha(\tau,s) =  K(\tau,s) k_\alpha(\tau-s)$. The resolvent $R_\alpha(\tau,s)$ inherits the weak singularity $(\tau-s)^{\alpha - 1}$ and can be represented via Neumann series.
Since each term in the series is uniformly bounded on $\calD$, the series converges absolutely and uniformly for all $\alpha \in (0,1)$. For more details, see \cite{ItkinLiptonMuraveyBook}.

In practice, by a change of variables $s \to \tau -\nu^2$, e.g. the second integral in \eqref{PDE_EB} transforms to
\begin{align} \label{tr2}
\int_0^\tau \eta(s, y(s)) \frac{y(\tau) - y(s)}{\tau-s} \frac{e^{-\frac{(y(\tau)-y(s))^2}{4(\tau-s)}}}{\sqrt{\pi(\tau-s)}} ds &= \frac{2}{\sqrt{\pi}} \int_0^{\sqrt{\tau}} \eta(\tau -\nu^2, y(\tau -\nu^2)) \frac{y(\tau) - y(\tau - \nu^2)}{\nu^2} \\
&\cdot e^{-\frac{\left(y(\tau )-y\left(\tau -\nu ^2\right)\right)^2}{4 \nu ^2}} d\nu. \nonumber
\end{align}

By a Taylor series expansion,
\begin{align}
\frac{y(\tau) - y(\tau - \nu^2)}{\nu^2} &= y'(\tau) + O(\nu^2), \\
\frac{\left(y(\tau) - y(\tau - \nu^2)\right)^2}{4 \nu^2} &= \frac{1}{4} \nu^2 y'(\tau)^2 - \frac{1}{4} \nu^4 \left(y'(\tau) y''(\tau)\right) + O(\nu^5), \nonumber
\end{align}
and the exponent in \eqref{tr2} vanishes as $\nu \to 0$. Consequently, the integral in \eqref{tr2} remains nonsingular.

Alternatively, this integral can be computed numerically using singularity subtraction. This method isolates the singular part for analytical solution, allowing the remaining regular part to be handled with standard numerical quadratures. In particular, given some function $g(s)$ we have
\begin{align} \label{tr21}
\int_0^{\tau} & g(s) \frac{e^{-\frac{(y(\tau)-y(s))^2}{4(\tau-s)}}}{\sqrt{\pi(\tau-s)}} ds =
\int_0^{\tau} \frac{g(s) e^{-\frac{(y(\tau)-y(s))^2}{4(\tau-s)}} - g(\tau)}{\sqrt{\pi(\tau-s)}} ds
+ g(\tau) \int_0^{\tau} \frac{1}{\sqrt{\pi(\tau-s)}} ds \\
&= \int_0^{\tau} \frac{g(s) e^{-\frac{(y(\tau)-y(s))^2}{4(\tau-s)}} - g(\tau)}{\sqrt{\pi(\tau-s)}} ds + 2 g(\tau) \sqrt{\tau/\pi}. \nonumber
\end{align}
The first integral in the last line vanishes at $s = \tau$. This can be shown by a Taylor series expansion of the integrand, and this vanishing behavior removes the singularity.

\subsection{The exercise boundary for the American Call option}

For the American Call option the EB can be computed in a way similar to that \cref{ebPut}. Indeed, again by
by a standard argument, \citep{ContVolchkova2005, klebaner2005} for the model in \cref{GBM-41,divDef}, the American Call option price $C(t,S)$ in the {\it continuation} region $S \in \calC = [0, S_B(t)]$ solves a parabolic PDE
\begin{equation} \label{PDE_ebC}
\fp{C}{t} + \dfrac{1}{2}\sigma^2(t) S^2 \sop{C}{S} +  [a(t) S - b(t)] \fp{C}{S} = r(t) C,
\end{equation}
subject to the boundary conditions
\begin{equation} \label{bcEBC}
C(t,S_B(t) = S_B(t) - K, \qquad C(t,0) = 0,
\end{equation}
and the terminal condition
\begin{equation} \label{tcEBC}
C(T,S) = (S - K)^+ = 0,
\end{equation}
since $S_B(T) = K$.

Again, by a change of variables similar to that in \eqref{trEB}, but now with
\begin{align} \label{trEBC}
S &\to \frac{K}{\alpha(t)}e^{-x}, \qquad C(t,x) = U(\tau,x)e^{\int_T^t r(k) dk}, \qquad (\tau, x) \in [0,\tau(0)]\times [y(\tau),\infty),
\end{align}
the PDE in \eqref{PDE_ebC} can be transformed to the PDE in \eqref{PDE_EB} with the moving left boundary $y(\tau)$
\begin{align} \label{PDE_uC}
\fp{U}{\tau} &= \sop{U}{x}  + \lambda(\tau, x), \qquad \lambda(\tau, x) = e^{x} \gamma(\tau) \fp{U}{x}, \\
U(0,x) &= - z(0,x), \qquad U(\tau,x)\Big|_{x \uparrow \infty} = U(\tau,y(\tau)) = 0, \qquad  z(\tau,x) = K \beta(\tau) \left( e^{-x} - \alpha(\tau)\right). \nonumber
\end{align}
This is the same problem as in \eqref{PDE_EB}, with change of signs in the definition of $z(\tau)$ and $\lambda(\tau,x)$. Consequently, $y(\tau)$ solves a system of equations \eqref{solPut} and \eqref{VolEB1} with the corresponding changes of signs. In particular, the integral $J_1$ and $\eta(s,\xi), \zeta(s,\xi)$ are now defined as
\begin{align}
J_1 &= -\int_{y(s)}^{\infty} e^{\xi} U(s,\xi) \Bigg\{- \frac{1}{2\sqrt{\pi(\tau - s)}}
\left[ e^{-\frac{(x-\xi )^2}{4 (\tau-s )}} - e^{-\frac{(\xi + x-2 y(\tau ) )^2}{4 (\tau-s )}} \right] \\
&+ \frac{1}{4\sqrt{\pi(\tau - s)^3}} \left[(x-\xi)e^{-\frac{(x-\xi )^2}{4 (\tau-s )}} + (x + \xi - 2y(\tau)) e^{-\frac{(\xi +x-2 y(\tau ))^2}{4 (\tau-s )}}  \right] \Bigg\} d\xi, \nonumber \\
\zeta(\tau,x) &= 2 \tau z(\tau,x) (x - y(\tau))e^{x} - K \beta(\tau), \nonumber \\
\eta(\tau,x) &= z(\tau,x) \left[ (x-y(\tau))^2 (1 + 4 \tau^2)  - 2 \tau y'(\tau) (x-y(\tau)) - (2 \tau + \bar{r}(\tau)) \right], \nonumber \\
&+ K \beta(\tau) e^{-x} \left[4 \tau (x-y(\tau)) +  \rho'(\tau)\right]. \nonumber
\end{align}
and \eqref{VolEB1} transforms to
\begin{align} \label{VolEB1c}
- \frac{1}{\sqrt{\pi \tau}} & \int_{y(0)}^\infty U'_\xi(0,\xi) e^{-\frac{(\xi - y(\tau))^2}{4\tau}} d\xi = \int_0^\tau \int_{y(s)}^\infty \frac{e^{-\frac{(\xi - y(\tau))^2}{4(\tau - s)}}}{2\sqrt{\pi(\tau - s)}} \frac{\xi - y(\tau)}{\tau-s}  \eta(s,\xi) e^{-s(\xi-y(s))^2} d\xi ds, \\
&+ \int_0^\tau \gamma(s) \int_{y(s)}^{\infty} \left[ e^{\xi} U'_\xi(s,\xi) - \zeta(s,\xi) e^{-s (\xi-y(s))^2} \right] \frac{\xi - y(\tau)}{2\sqrt{\pi (\tau - s)^3}} e^{-\frac{(\xi - y(\tau))^2}{4(\tau-s)}} d\xi ds, \nonumber
\end{align}
with
\begin{align}
J_0 &= - \frac{1}{\sqrt{\pi \tau}} \int_{y(0)}^\infty U'_\xi(0,\xi) e^{-\frac{(\xi - y(\tau))^2}{4\tau}} d\xi =
- K e^{\tau - y(\tau)} \erfc\left(\frac{2 \tau - y(\tau) }{2 \sqrt{\tau }}\right).
\end{align}

Also, due to the definitions in \eqref{trEBC}, now we have
\begin{align} \label{b3}
\Psi(\tau, y(\tau)) &= 0, \qquad y(\tau) = \tau - \rho(\tau) - \log(S_B(\tau)/K),
\end{align}
Hence, for a Call option, $y(\tau)$ is a concave function, implying $y'(s) < 0$ and $y(s) < 0$.

Same as for the American Put option, in case of no dividends (i.e., $\gamma(t) = q(t) = 0$), \eqref{VolEB1} transforms to \eqref{VolEB-a1}. But, since for the American Call option $S_B(s) \ge K$ and $S_B'(s) > 0$,  this implies at $r(s) > 0$,
\begin{align} \label{yDifPos}
y(\tau) - y(s) = (\tau-s)\left(1 + \frac{1}{\tau-s}\int_s^\tau \frac{2 r(k)}{\sigma^2(k)} \right) - \log\left( \frac{S_B(\tau)}{S_B(s)} \right) < 0,
\end{align}
if $\tau > s$.

Thus, the sign of the first integral in the right-hands part of \eqref{VolEB1c} coincides with that of $\eta(s,\xi)$. This statement holds because $\xi - y(\tau) > 0$ even when $\xi = y(s)$. Furthermore, the integrand decays rapidly for large $\xi$ due to the product of two Gaussian exponential terms. Consequently, the integral's dominant contribution comes from the region near the lower limit, $\xi = y(s)$. In this limit we have
\begin{align} \label{pAtY1}
\eta(s, y(s)) &= - (2 s + \br'(s)) z(s, y(s)) + K \beta(s) e^{-y(s)} \rho'(s) = - 2 s z(s, y(s) + K \beta(s) e^{\bar{r}(s)} \bar{r}'(s), \\
z(s, y(s)) &= K e^{\bar{r}(s)} \left(\frac{S_B(s)}{K} -1 \right)  \ge 0. \nonumber
\end{align}
This result holds because for no dividends $\bar{r}(s) = \rho(s)$. Thus, $\eta(s,y(s)) \geq 0$ (at least, for small $s$).
As the result, the right-hand side of \eqref{VolEB1c} is non-negative, while the left-hand side is non-positive. Therefore, the integral equation admits no real-valued solutions.

A more delicate analysis reveals that in this case the only correct solution would be $y(\tau) = -\infty$ or $S_B(\tau) = \infty$. Indeed, in this case the right-hands part of \eqref{VolEB1c} tends to 0, so does the left-hands part due to the identity
\begin{equation}
\lim_{y(\tau) \to - \infty } e^{\tau - y(\tau)} \erfc\left(\frac{2 \tau - y(\tau) }{2 \sqrt{\tau }}\right) = 0.
\end{equation}

This means that the EB for an American Call doesn't exist (or, it moves to infinity) for $r > 0$ and the whole domain $S \in [0, \infty)$ is just a continuation region. To recall,  although this fact is well-established, it is typically argued from a financial perspective, while here we derive it mathematically through an analysis of \eqref{VolEB1c}. An absence of the EB implies that early exercise is never optimal for the American Call option. Consequently, its price equals that of the corresponding European option, and the EEP is zero.

However, when $r(s) < 0$, the derivative $S'_B(\tau)$ becomes negative. This occurs because it is beneficial to exercise early to avoid paying a higher effective strike in the future. In other words, if you exercise early, you get $S-K$ immediately. If you wait, the option value may increase, but the cash $K$ you would pay in the future is actually becoming less valuable (since $r(s) < 0$, the present value of $K$ increases over time). This means that the strike price
$K$ becomes more expensive in real terms as time passes.

Consequently, the right-hands part of \eqref{VolEB1c} will now contain negative terms implying this integral equation admits a real-valued solution $y(\tau) < 0$, a result also supported by its numerical solution.

\section{De-Americanization} \label{sect_deAmer}

The de-Americanization of American options refers to the process of converting an American option, exercisable at any time before expiration, into an equivalent (in the implied volatility) European option, which can only be exercised at maturity. This transformation simplifies pricing and analysis, particularly when restoring local volatility from market American option prices $P_M(t,S)$ by using Dupire’s formula, \cite{Dupire:94}. For the model in \eqref{GBM-41}, this involves computing the volatility $\sigma(t)$ of the underlying asset’s dynamics so the theoretical (model dependent) price coincides with the market option.

Traditionally, numerical methods such as binomial or trinomial trees are employed to determine this implied volatility $\sigma$ (which is assumed to be constant), see \cite{Burkovska2018} and reference therein, while a semi-analytical approach of \cite{Henry-Labordere2017} provides more tractability despite less accurate. However, when discrete dividends are present, de-Americanization becomes more complex. At each ex-dividend date, the stock price must be adjusted by the dividend amount, and the option value must be computed under the no-early-exercise assumption, i.e., as a European-style continuation value, see \cite{Vellekoop2006,ArealRodriges2013} among others.

It turns out that the approach proposed in this paper can also be naturally applied to de-Americanization. As detailed in Appendix~\ref{app1}, the algorithm for computing the transition density for the model with discrete dividends (given in \cref{GBM-41,divDef}) can be adapted to calculate the theoretical price $P(t,S)$ of the American option. This allows solving the equation
\begin{equation} \label{ivEq}
P(t, S) = P_M(t, S),
\end{equation}
iteratively to determine $\sigma(t)$.

Once this equation is solved and the implied volatility is determined, the corresponding European option price can be obtained by evaluating the first expectation in \eqref{decompGen} - that is by computing the integral of the discounted payoff multiplied by the already known transition density. Notably, this value is already computed at every iteration while solving for $\sigma(t)$.

Furthermore, note that this result depends on $\sigma(t)$ only through the definition of $\tau$ in \eqref{tau}. Introducing the mean volatility $\Sigma$ as
\begin{equation} \label{tauS}
\Sigma^2 = \frac{1}{2 T} \int_t^T \sigma^2(k) \, dk \ge 0, \qquad \text{so that} \quad \tau = \Sigma^2 T,
\end{equation}
we can determine it by solving the algebraic equation \eqref{ivEq}. At each iteration step for $\Sigma$, the computation of $P(\tau, x)$ follows the algorithm in \eqref{procP}, which requires evaluating $2(N_d + N_p)$ integrals with Gaussian kernels. Consequently, the computational complexity of this method is $O(4 (N_d + N_p) I N_x)$, where $I$ is the number of iterations needed to achieve convergence within the desired tolerance, and $N_x$ is the number of points in $x$-space used to compute the integrals. This can be compared with the complexity of the binomial tree method which is $O(N_x M^2)$ with $M$ the number of time steps, and $M \gg N_d + N_p$ since the temporal grid must encompass all ex-dividend dates as well as additional time steps to ensure sufficient accuracy (for the binomial tree method, the accuracy is only $O(\Delta t)$, corresponding to a first-order temporal approximation). Moreover, in the case of discrete cash dividends the binomial tree loses its recombing property.

When using the finite-difference (FD) method, a standard approach for valuing American options is: a) approximating the American option as a Bermudan option with a large number of exercise opportunities, and b) applying Richardson extrapolation to a series of Bermudan options with increasing exercise dates. However, this method introduces errors due to the discrete exercise schedule. Furthermore, the Bermudan option price is computed iteratively over time with some step size $\Delta t = T/M$. At each time step $t_m$, the option value is given by
\begin{equation}
P(t_m, x) = \max (E(t_m, x), P_E(t_m, x)),
\end{equation}
\noindent where $P_E(t_m,x)$ is the continuation value (a European option price) and $E(t_m,x)$ is the exercise value. While this representation suffices for pricing, it suffers from poor accuracy when computing option Greeks, especially near the early exercise boundary (EB). To mitigate this issue, a penalty method is recommended \cite{Zvan1998,Halluin2004}, although it makes the whole algorithm slower. Overall, the complexity of the FD method is $O(N_x M I_p)$, with $I_p$ being the number of iterations in the penalty method.

Thus, our method should be faster due to: a) the fact that $M \gg N_d + N_p$ because between the ex-dividend dates we employ a semi-analytical solution of the problem, so we don't need a dense grid, and b) computation of spatial integrals is done using the FGT where the error decays exponentially when increasing $p$ – the degree at which a Hermite expansion of the Gaussian kernel is truncated. Usually, moderate values of $p \approx 10$ provide an error near machine precision, while the FD method spatial error is normally $O(h^2)$ with $h$ being the spatial step. Thus, when comparing with the FD method, the FGT can deliver the same error while running with lower $p$, providing even faster performance.

Alternatively, rather than solving an integral equation for the transition density, one can directly use the corresponding option pricing equation (e.g., \eqref{Volt2}) to infer the implied time-dependent volatility $\sigma(t)$.

\subsection{De-Americanization via implied strikes}

Similar to the concept of implied volatility, we can introduce the alternative notion of an {\it implied strike} $\mathcal{K}$. Given a maturity $T$ and a volatility function $\sigma(t)$, the implied strike is defined as the solution to \eqref{ivEq}. To the best of our knowledge, this idea was first proposed by A.~Skabelin in \cite{Skabelin2015}. Under this approach the implied volatility could be chosen as an input of the model, for instance being equal to the constant ATM implied volatility.

In this framework, de-Americanization means that instead of working with implied volatilities, we consider a set of implied strikes $\mathcal{K}$ that make the theoretical price of an American option match its market price. By assuming that $\mathcal{K}$ remains the same for both American and European options, we can obtain a de-Americanized European option price.
This adjusted price can then be used to recover the local volatility from American option market prices $P_M(t,S)$ via Dupire’s formula \cite{Dupire:94}, albeit with a slight modification. Under this new framework, the option price is expressed as a function of the implied strike $\calK$ rather than the strike $K$, where $\calK$ is defined in terms of the variable $x$ by \eqref{tr1}, i.e.,
\begin{equation}
\calK = \alpha(0) S e^{-x}.
\end{equation}
Thus, without ambiguity, $x$ can also be interpreted as a dimensionless implied strike.

Recall that the values of $x$ are obtained by solving \eqref{ivEq}. Then, in the Dupire formula, for instance, for the Call option, the local volatility is given by:
\begin{equation}
\sigma^2(K,T) = \frac{\fp{C}{T} + [(r(T)-q(T)] K \fp{C}{K} + q(T) C}{\frac{1}{2} K^2 \sop{C}{K}}.
\end{equation}
However, now the following substitutions must be applied:
\begin{align}
C(\tau,x) &= W(\tau, x) e^{-\int_0^\tau r(k) dk}, \quad \fp{C}{T} \to \fp{C}{\tau} \fp{\tau}{T} + \fp{C}{x} \fp{x}{T}, \\
\fp{C}{K} &\to \fp{C}{x} \fp{x}{K}, \qquad \sop{C}{K} \to \fp{C}{x} \sop{x}{K} + \sop{C}{x} \left(\fp{x}{K}\right)^2,
\end{align}
where
\begin{align}
\fp{x}{K} &= -\frac{1}{K}, \quad \sop{x}{K} = \frac{1}{K^2}, \quad \fp{x}{T} = -\frac{1}{2} \sigma^2(T) + a(T), \quad \fp{\tau}{T} = \frac{1}{2}\sigma^2(T),
\end{align}
and the function $\tau(t,T)$ is defined in \eqref{tau}.

This approach provides a notable computational advantage because the implied strikes $\mathcal{K}$ (or $x$) can be calculated simultaneously for all market quotes across different strikes $K$ (for a fixed maturity $T$). Indeed, \eqref{Volt2} can be made "autonomous", i.e., independent on $K$ since by the definitions of $W(\tau, x)$ and $\Psi(\nu, \xi)$ in \cref{tr1,PDE1}, both terms are proportional to $K$, meaning the entire equation \eqref{Volt2} can be divided by $K$ to eliminate explicit strike dependence. As the variable $x$ itself depends on $K$, the solution to \eqref{Volt2} should be determined for a range of $x$ values corresponding to market quotes. However, thanks to the FGT technique, this computation can be performed in a single sweep, as the method yields simultaneous outputs for multiple $x$. Once \eqref{Volt2} is solved across the relevant $x$ values which are obtained using the given spot price $S$ and market strikes $K$, the implied strikes $\mathcal{K}$ can be reconstructed (e.g., via interpolation) from the definition of $x$ in \eqref{tr1}.

Now, what happens when the option maturity $T$ changes? According to the definition of $\tau$ in \eqref{tau}, we can adjust the volatility function $\sigma(t)$ to maintain the original value of $\tau$. Consequently, while \eqref{Volt2} remains unchanged, the market quotes $P_M(t,S)$ on the left-hand side of \eqref{ivEq} (or \eqref{Volt2}) will vary.

This means we must solve the same equation but with different left-hand sides. Importantly, this implies that solutions for all maturities $T$ can be obtained simultaneously by considering multiple left-hand sides, while the integrals in the right-hand side need only be computed once per iteration. As a result, all implied strikes $\mathcal{K}$ for various combinations of $T$ and $K$ can be determined in a single computational pass.

This approach is conceptually similar to the method proposed in \cite{CarrItkinStoikov2020} (see also \cite{Stefanica2020}). The authors note that while the traditional method for computing Black-Scholes implied volatility is relatively straightforward, it faces significant challenges when applied to a wide range of model (or contract) parameters simultaneously. Specifically, iterative root-finding algorithms may converge slowly due to two key issues: (i) the lack of a universally good initial guess to initiate the iterations, and (ii) the potential non-existence of solutions or extreme numerical sensitivity in certain parameter regions where the model price exhibits low sensitivity to changes in implied volatility.

A counterintuitive yet promising alternative is to replace the algebraic equation like \eqref{ivEq} with a more complex object, such as a PDE. At first glance, this substitution might seem impractical, as PDEs, particularly nonlinear ones, require specialized numerical methods that are far more involved than simple root-finding algorithms. However, the PDE framework offers a crucial advantage: whereas the algebraic equation must be solved independently at each point in the parameter space (e.g., for every strike and maturity), the PDE approach enables simultaneous solutions across multiple points via time-marching sweeps. Moreover, linear PDEs can be solved non-iteratively, eliminating the need for an accurate initial guess altogether.

A potential drawback of using implied strikes instead of implied volatilities is the need to construct an implied volatility surface from market quotes. If high precision is not essential, this can be achieved through various approximations, since the local volatility function is already known. For example, the approximation proposed in \cite{Berestycki2002} can be applied for this purpose.

\section{Numerical examples} \label{numExp}

This section presents numerical examples illustrating the methodology from the previous sections. The corresponding MATLAB code can be downloaded from \href{ https://github.com/itkinal/AmericanOptionsDividends/archive/refs/tags/1.1.zip}{this GitHub repository}.

The primary objective of this section is the computation of the EB. The American option price could be subsequently found using this EB. The procedure involves calculating the process's transition density (as outlined in \cref{trDensity}) and then substituting both the density and the EB into equation \eqref{decompGen2}. Evaluating the resulting integrals provides the European option value and the EEP.

In terms of computational cost, solving \eqref{PDE2} for the transition density is far less intensive than finding the EB, by a factor of 10 to 12. This disparity arises because the density solution uses a direct marching method in time, while the EB computation relies on an iterative root solver. Therefore, the total computation time is dominated by the EB, and the cost of obtaining the transition density and two remaining integrals is negligible in comparison.

\subsection{The EB of an American Put on a non-dividend-paying asset}

The first test case computes the EB for an American Put option on a non-dividend-paying asset, whose dynamics follows a time-homogeneous GBM. The results are then compared to those obtained from the binomial tree method, \cite{Seydel2017}.

The computation of EB via the GIT method involves numerically solving \eqref{VolEB-a1}. However, the integral in $\xi$ has a closed-form solution expressed in terms of special functions (via $\erf$). Therefore, we use this representation derived in
Appendix~\ref{appGA}
\begin{align} \label{altI1}
I &= \int_0^\tau \eta(s,y(s)) \frac{e^{-\frac{(y(s) - y(\tau))^2}{4(\tau - s)}}}{\sqrt{\pi(\tau - s)}} ds + I_2, \\
I_2 &= \int_0^\tau \int_{y(s)}^\infty \frac{\eta'_{\xi}(s,\xi) - 2 s \eta(s,\xi)(\xi-y(s))}{\sqrt{\pi(\tau - s)}} e^{-\frac{(\xi - y(\tau))^2}{4(\tau - s)} - s(\xi-y(s))^2} d\xi ds \nonumber\\
&= \int_0^\tau \frac{e^C}{2\sqrt{\pi(\tau - s)}} \left[ J_{1,1} + e^{-y(s)} J_{1,2} \right] ds, \nonumber
\end{align}
so, \eqref{VolEB-a1} now reads
\begin{align} \label{VolEB-a2}
- K e^{\tau + y(\tau)} \left[ 1 + \erf\left( \frac{2\tau + y(\tau)}{2 \sqrt{\tau}}\right) \right] =
\int_0^\tau \frac{ds}{\sqrt{\pi(\tau - s)}} \Big\{ \eta(s,y(s)) e^{-\frac{(y(s) - y(\tau))^2}{4(\tau - s)}}
+ \frac{e^C}{2} \left[ J_{1,1} + e^{-y(s)} J_{1,2} \right] \Big\},
\end{align}
where integrals $J_{1,1}, J_{1,2}$ and function $C$ are defined in Appendix~\ref{appGA}.

As this is discussed in detail in \cref{weakSing}, the first integral is an integral with weak singularity. Accordingly, based on \cref{weakSing}, we re-write it as
\begin{align*}
\int_0^\tau \eta(s,y(s)) \frac{e^{-\frac{(y(s) - y(\tau))^2}{4(\tau - s)}}}{\sqrt{\pi(\tau - s)}} ds
= \int_0^\tau \frac{ds}{\sqrt{\pi (\tau - s)}} \left[ \eta(s,y(s)) e^{-\frac{(y(s) - y(\tau))^2}{4(\tau - s)}} - \eta(\tau, y(\tau)) \right] + 2 \eta(\tau,y(\tau)) \sqrt{\frac{\tau}{\pi}}.
\end{align*}
Using a Taylor series expansion around $s \to \tau$, it can be shown that the integrand in the right-hands part vanishes at $s \to \tau$, implying that the whole integral in the left-hands part has no singularities in this form.

Regarding the second integral in \eqref{VolEB-a2}, which is discussed in Appendix~\ref{appGA}, it has no singularity at $s \to \tau$. The contribution of this integral is minor compared to the first one at $s \to \tau$. The same holds at $s \to 0$ if $\rho'(0) \sim O(1)$. However, this hierarchy at $s \to 0$ reverses for small $\rho'(0)$, causing the second term to dominate. Such small values of $\rho'(s)$ are typically observed in environments characterized by high volatility and low instantaneous interest rates.

Without loss of generality, we model the time-dependent coefficients of the GBM model as
\begin{equation} \label{ex}
r(t) = r_0 e^{-r_k t} + r_1, \qquad q(t) = q_0 e^{-q_k t} + q_1, \qquad \sigma(t) = \sigma_0 e^{-\sigma_k t} + \sigma_1,
\end{equation}
\noindent where $r_0, r_1, r_k, q_0, q_1, q_k, \sigma_0, \sigma_1, \sigma_k$ are constants. In the first example the values of these parameters are given in Table~\ref{tab1}, i.e. first, we explore a time-homogeneous GBM model with no dividends.
\begin{table}[!htb]
\begin{center}
\begin{tabular}{|c|c|c|c|c|c|c|c|c|}
\hline
$r_0$ & $r_1$ & $r_k$ & $q_0$ & $q_1$ & $q_k$ & $\sigma_0$ & $\sigma_1$ & $\sigma_k$ \\
\hline
0.01 & 0.0 & 0.0 & 0.0 & 0.0 & 0.0 & 0.6 & 0.0 & 0.0 \\
\hline
\end{tabular}
\caption{Model parameters for the numerical test 1.}
\label{tab1}
\end{center}
\end{table}
\begin{figure}[!htp]
\begin{center}
\hspace*{-0.3in}
\subfloat[]{\includegraphics[width=0.5\textwidth]{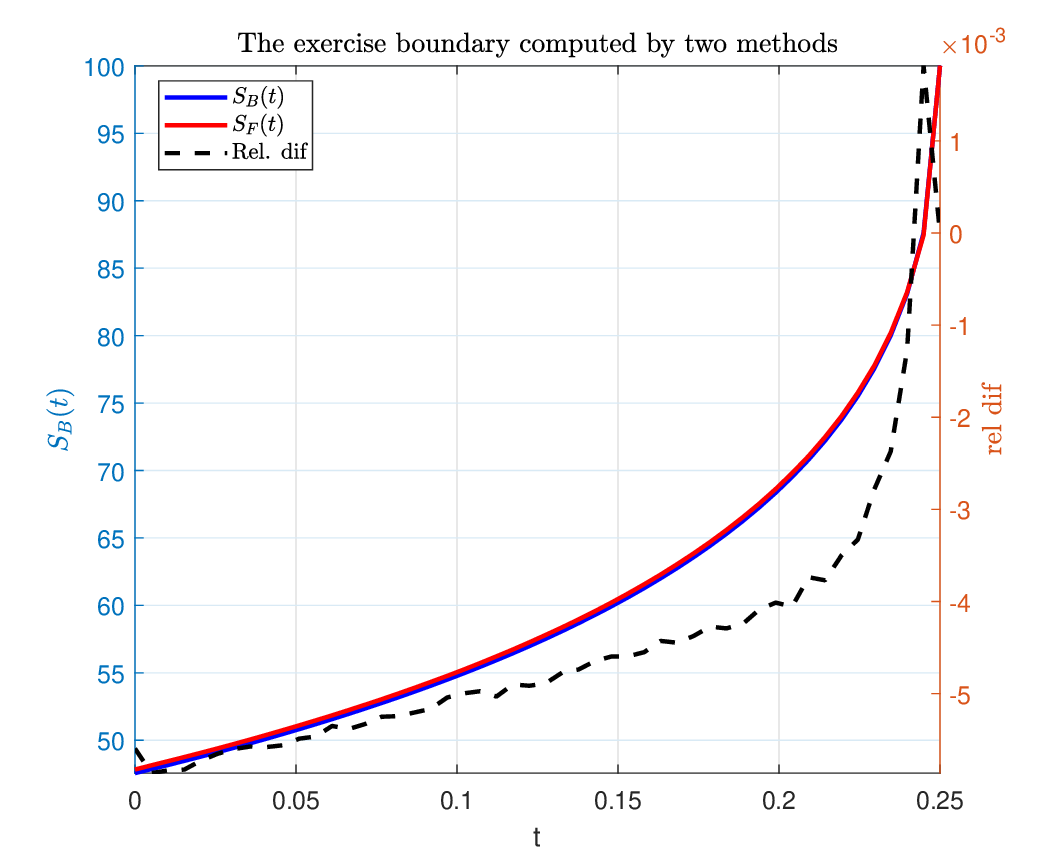}}
\subfloat[]{\includegraphics[width=0.5\textwidth]{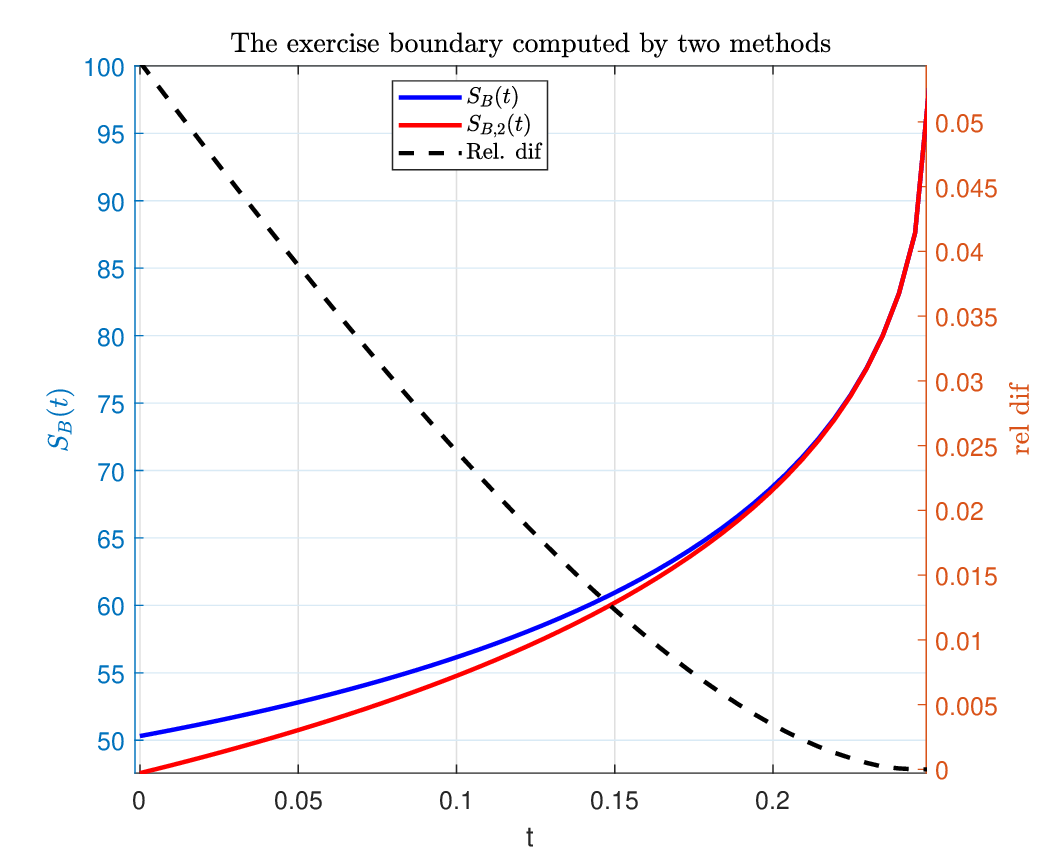}}
\end{center}
\caption{Early exercise boundaries for an American Put option under the time-homogeneous GBM model: (a) Using the model parameters from Table~\ref{tab1} with no dividends. The boundary $S_B(t)$ is computed using our method, while $S_F(t)$ is from the binomial tree method. (b) The same test where the EB is computed by solving \eqref{VolEB-a2} taking into account only the first term under the integral (curve $S_B(t)$) and both terms (curve $S_{B,2}(t)$).}
\label{Test1}
\end{figure}

The Volterra equation in \eqref{VolEB-a2} is solved using the trapezoidal rule. We use the previously computed value $y(t_{i1})$ as an initial guess for the iterative solver, which typically converges within 10-12 iterations to reach the tolerance \num{1.0e-8}.
\footnote{
    Higher-order quadrature methods, such as Simpson’s rule (which achieves $O((\Delta t)^4)$ accuracy) or adaptive schemes, could further enhance precision. These improvements require minimal implementation changes and no fundamental algorithmic modifications. In contrast, achieving similar accuracy gains with finite difference, tree, or Least Squares Monte Carlo (LSMC) methods remains challenging.
} All computations were performed in MATLAB on a system with two Intel Quad-Core i7-4790 CPUs at 3.80 GHz.

The first numerical test is conducted using the parameters $T = 0.25, K = 100$, and $S = 100$. In Fig.~\ref{Test1}(a) the EB for the American Put option is plotted as a function of time $t$. Our method computed this boundary in 0.05 seconds using only $N = 50$ time steps. This small number of steps proved to be sufficient, as the output is stable and remains practically unchanged with further refinement of the time grid. This is due to the fact that the trapezoid quadratures provide a second order approximation $O(\Delta \tau)$.

\begin{figure}[!htp]
\begin{center}
\hspace*{-0.3in}
\subfloat[]{\includegraphics[width=0.5\textwidth]{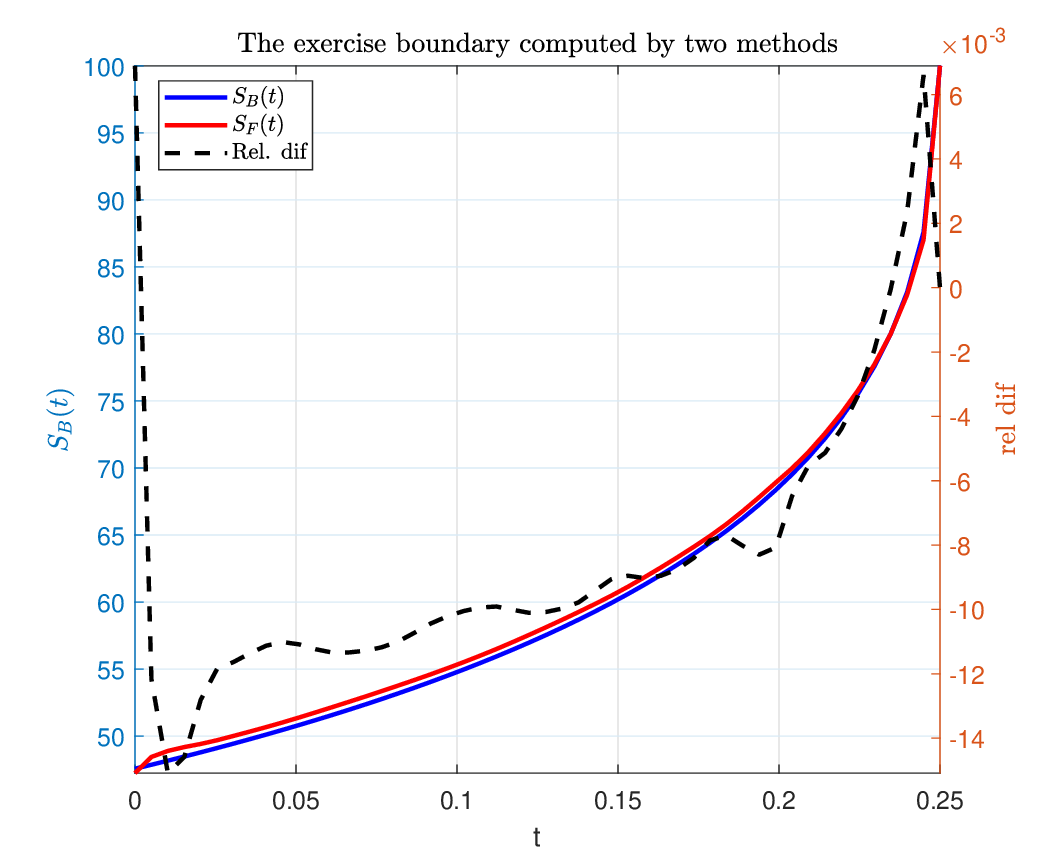}}
\subfloat[]{\includegraphics[width=0.5\textwidth]{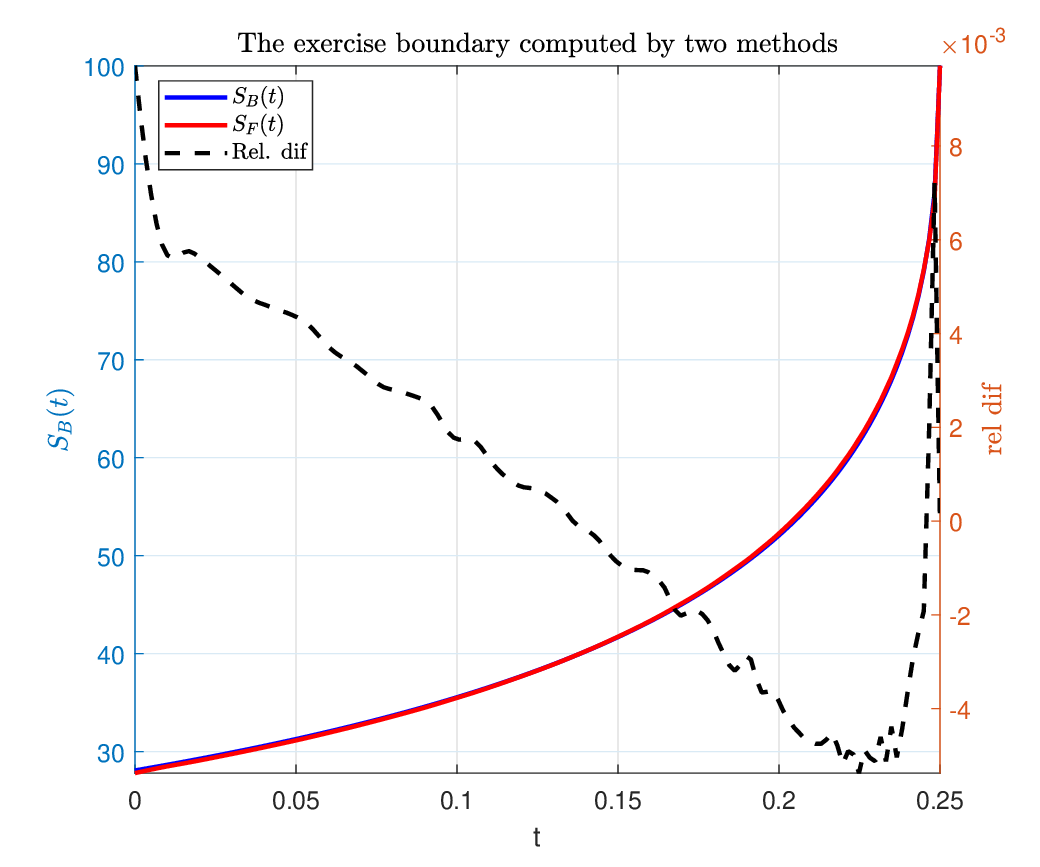}}
\end{center}
\caption{Early exercise boundaries for an American Put option under the time-homogeneous GBM model: (a) Using the model parameters from Table~\ref{tab1} with no dividends. The boundary $S_B(t)$ is computed using our method, while $S_F(t)$ is from the binomial tree method. (b) The same test is repeated with $r = 0.01, \sigma = 0.3$, and a coarse $400 \times 400$ tree grid, which produces a non-monotonic early exercise boundary.}
\label{Test1-sup}
\end{figure}

Following the discussion in Appendix~\ref{appGA}, Fig.~\ref{Test1}(b) assesses the contribution of the two terms in the integral in \eqref{VolEB-a2}. This is done by computing the EB using only the first term (curve $S_B(t)$ ) and then using both terms (curve $S_{B,2}(t)$ ). The two curves are nearly identical close to maturity but deviate as one moves away from it. This behavior, however, is parameter-dependent. For instance, with $r=0.05$ and the other parameters from Table~\ref{tab1}, the deviation remains relatively small for all $t$ .

We compare our results against the EB computed using the binomial tree method \cite{Seydel2017}. With a standard $400\times 400$ grid, the binomial method runs in 0.03 seconds but yields a large error and a non-monotonic boundary for large $\tau$, as shown in Fig.~\ref{Test1-sup}(a). To eliminate this non-monotonicity (Fig.~\ref{Test1-sup}(b)), the grid must be increased to $1600 \times 3000$ points, which increases the runtime to 0.37 seconds.

Both methods are sensitive to the value of volatility. An increase in $\sigma$ requires a larger $N$ in our method and a larger grid size for the binomial tree. However, ensuring the stability of the binomial tree is non-trivial, as it requires manually adjusting the number of space (decreasing) or time steps (increasing) to avoid non-monotonicity and a loss of accuracy, with no automated algorithm to guide this choice. This is demonstrated in Fig.~\ref{Test1-sup}(b) for the parameters $r=0.01, \sigma = 1.0$. To achieve agreement between the outputs, we increased $N$ to 150 for our method and the binomial grid to $500\times 2000$. Consequently, the elapsed time increased to 0.10 seconds for our method and 0.12 seconds for the binomial tree (vs 0.04 secs for the grid of size $400\times 400$).

\subsection{The EB of an American Put with continuous dividends.}

\begin{figure}[!htp]
\begin{center}
\hspace*{-0.3in}
\subfloat[]{\includegraphics[width=0.5\textwidth]{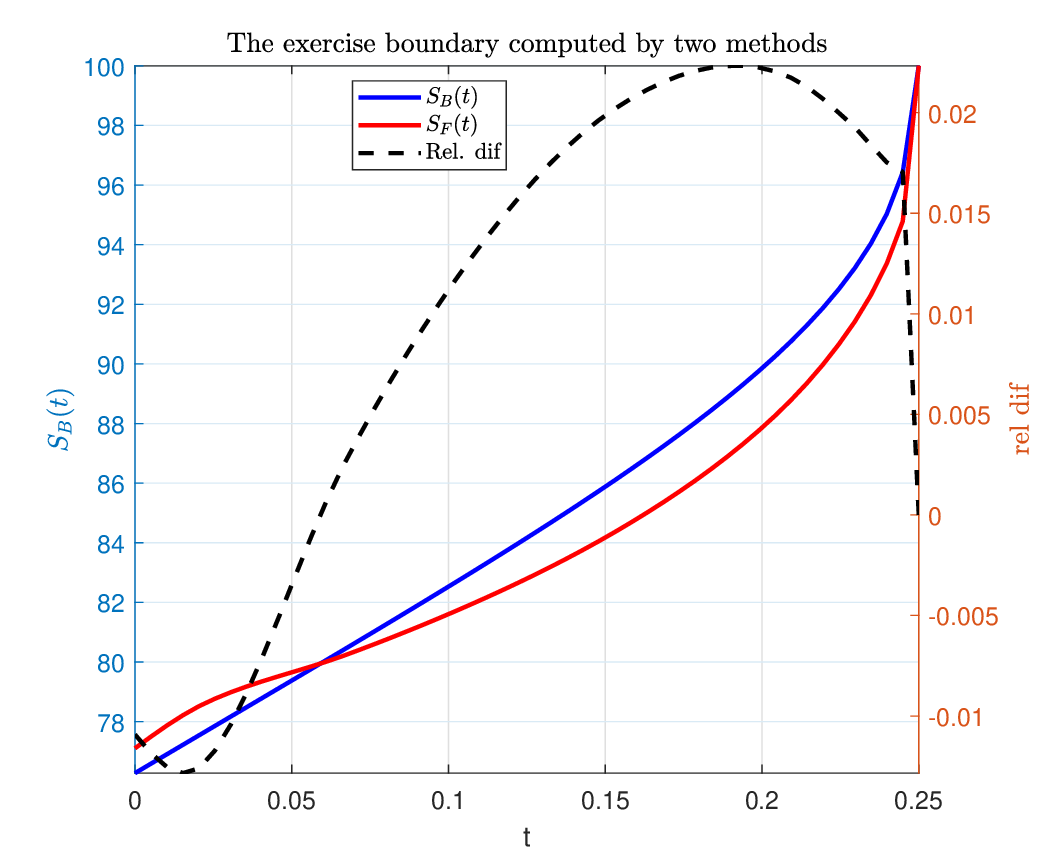}}
\subfloat[]{\includegraphics[width=0.5\textwidth]{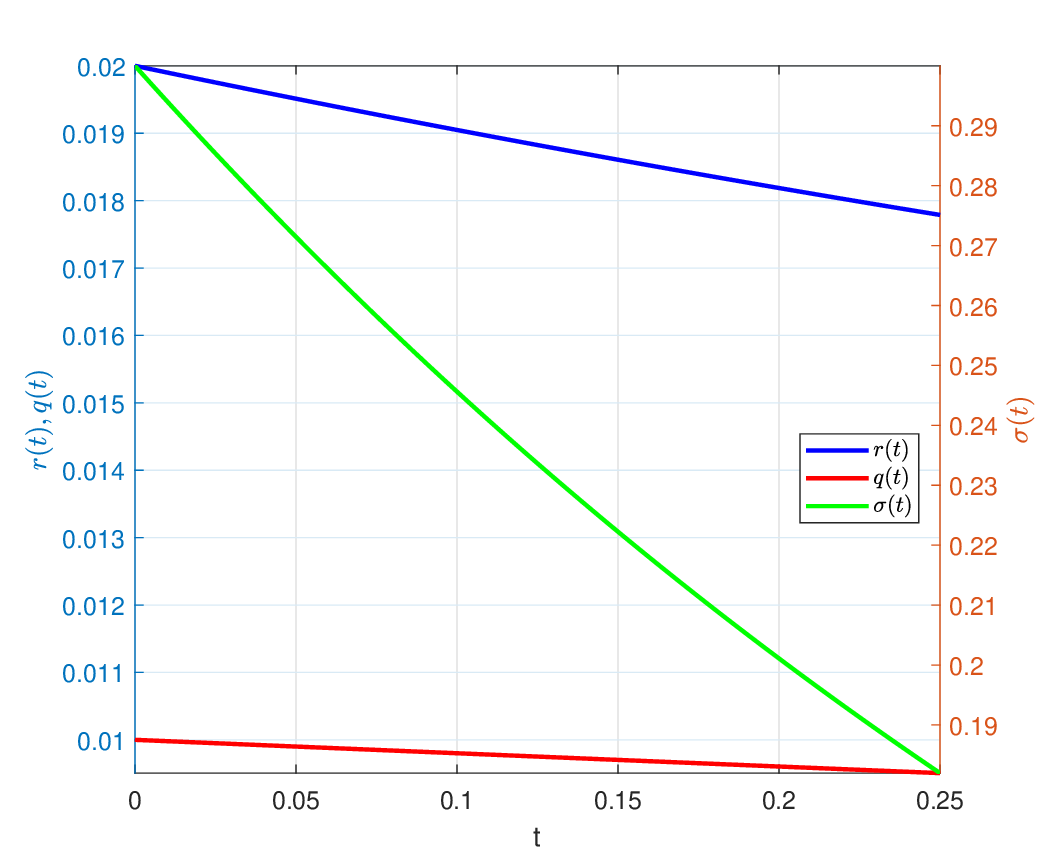}}
\end{center}
\caption{Early exercise boundaries for an American Put option under the time-inhomogeneous GBM model: (a) Using the model parameters from Table~\ref{tab2}. The boundary $S_B(t)$ is computed using our method, while $S_F(t)$ is from the binomial tree method with $500 \times 2000$ points. (b) Time-dependent parameters of the model.}
\label{Test2}
\end{figure}

To investigate the influence of continuous dividends, we modify our model in two ways: a) by including continuous dividends, and b) by making all model parameters time-dependent. These parameters are given in Table~\ref{tab2}.
\begin{table}[!htb]
\begin{center}
\begin{tabular}{|c|c|c|c|c|c|c|c|c|}
\hline
$r_0$ & $r_1$ & $r_k$ & $q_0$ & $q_1$ & $q_k$ & $\sigma_0$ & $\sigma_1$ & $\sigma_k$ \\
\hline
0.01 & 0.01 & 1.0 & 0.02 & -0.01 & 0.1 & 0.3 & 0.0 & 2.0 \\
\hline
\end{tabular}
\caption{Model parameters for the numerical test 2 (continuous dividends and time-inhomogeneous parameters.}
\label{tab2}
\end{center}
\end{table}
We compare our results with those from the binomial tree method, which uses average parameters defined as:
\begin{align}
r_a = \frac{1}{T} \int_0^T r(k) dk, \qquad q_a = \frac{1}{T} \int_0^T q(k) dk, \qquad
\sigma^2_a = \frac{1}{T} \int_0^T \sigma^2(k) dk,
\end{align}
where $r(t), q(t), \sigma(t)$ are given in \eqref{ex}. The results are shown in Fig.~\ref{Test2}.

The results indicate that even with a fine grid of $500\times 2000$, the binomial tree method produces non-monotonicity for large $\tau$. Furthermore, averaging the time-dependent parameters is ineffective, as the resulting EB diverges significantly from that computed by our time-inhomogeneous method ($N=50$) as shown in Fig.~\ref{Test2}(b). Our method is also computationally more efficient, with an elapsed time of 0.06 seconds compared to 0.08 seconds for the binomial tree. While efficient FD methods could resolve these issues with trees, our approach provides a clear advantage since it doesn't require a 2D space-time grid.

\subsection{The EB of an American Put with discrete proportional dividends} \label{secPropDiv}

We now examine the case of discrete proportional dividends, which are incorporated into the model via the function $a(t)$  in \eqref{divDef}. Since $a(t)$  only affects the function $\rho(\tau)$ , the inclusion of discrete dividends solely modifies this term. We consider $M$  dividends with amounts $d_i \ne 0$ paid at ex-dividend dates $t_{i}$.

\begin{figure}[!htp]
\begin{center}
\hspace*{-0.3in}
\subfloat[]{\includegraphics[width=0.5\textwidth]{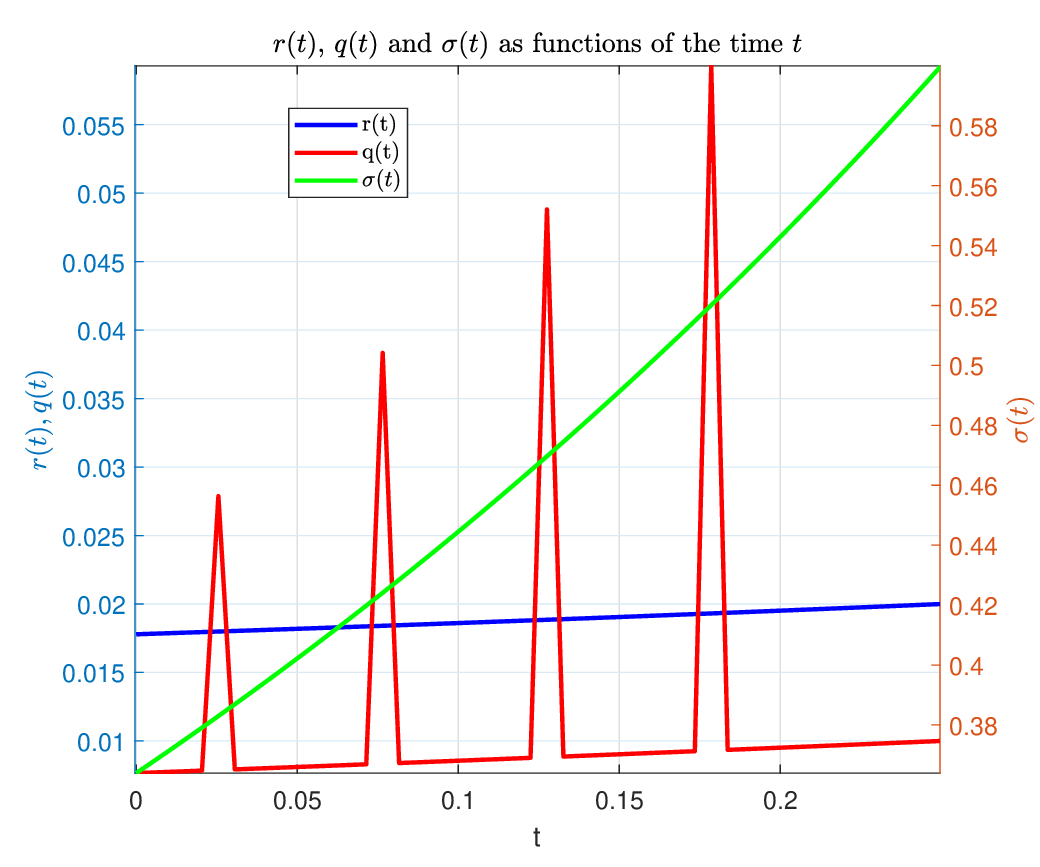}}
\subfloat[]{\includegraphics[width=0.5\textwidth]{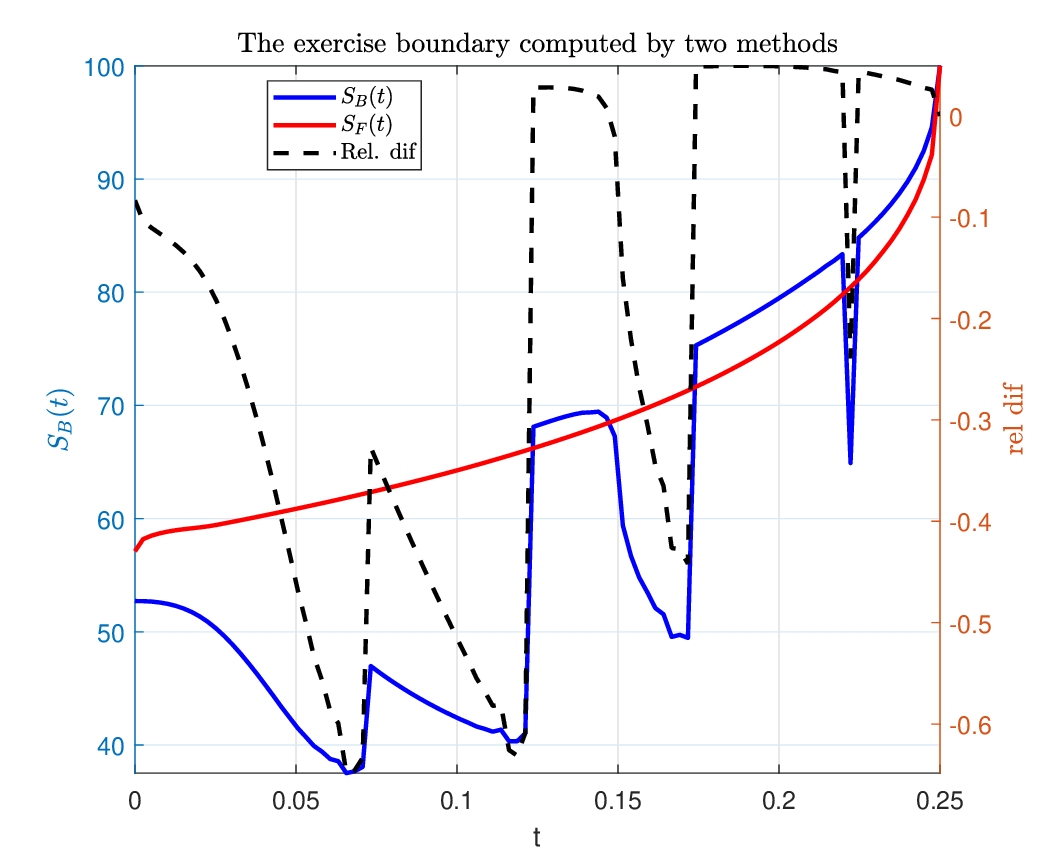}}
\end{center}
\caption{Early exercise boundaries for an American Put option under the time-inhomogeneous GBM model. (a) Time-dependent model parameters $r(t)$, $q(t)$, and $\sigma(t)$, accounting for discrete proportional dividends. (b) The EB $S_B(t)$ is computed using our method, while $S_F(t)$ is from a binomial tree method that uses no discrete dividends and constant parameters, obtained by averaging the time-dependent parameters over the option's lifetime.}
\label{divProp}
\end{figure}

We conduct a test with four ex-dividend dates at $t = [0.07, 0.12, 0.17, 0.22]$  and a maturity of $T=0.25$. The associated dividend amounts are $[5\%, 4\%, 3\%, 2\%]$. All remaining parameters are identical to those specified in \Cref{tab1,tab2}. The corresponded $r(t), q(t), \sigma(t)$ are shown in Fig.~\ref{divProp}(a).

A temporal grid resolution of $N=100$  is employed to accurately capture the ex-dividend dates, which are required to lie on the grid to preclude the need for interpolation. The results are presented in Fig.~\ref{divProp}(b). A comparative benchmark, shown by the red curve, is generated using a binomial tree with no discrete dividends. In this benchmark, the time-dependent continuous dividends are replaced by a constant continuous dividend yield, obtained by averaging the time-dependent dividends over the option's lifetime.

The elapsed computation time is almost unchanged at 0.08 seconds, as the algorithm's structure is preserved. The sole modification is the pre-computation of the function $a(t)$.

\begin{figure}[!htp]
\begin{center}
\hspace*{-0.3in}
\subfloat[]{\includegraphics[width=0.5\textwidth]{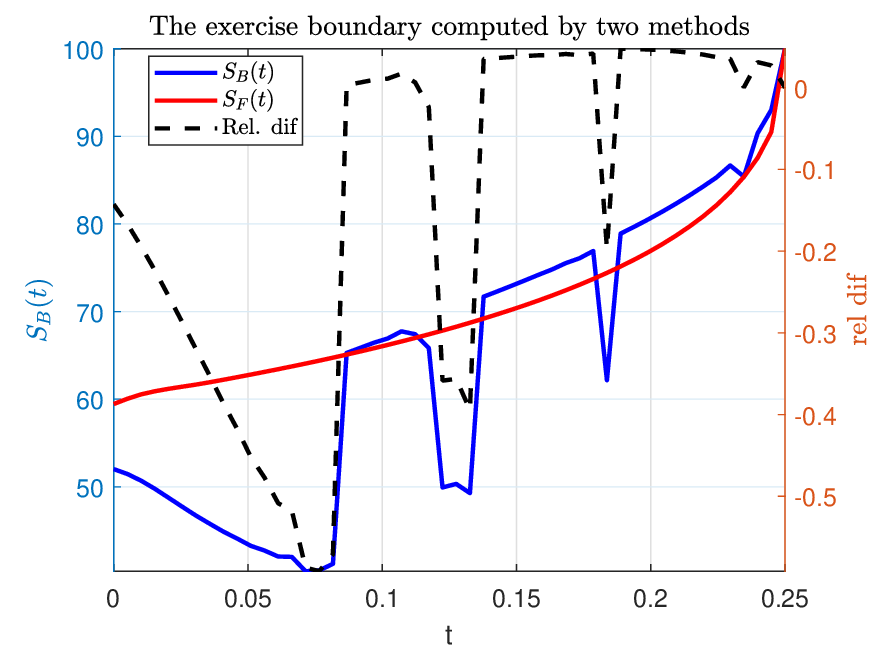}}
\subfloat[]{\includegraphics[width=0.5\textwidth]{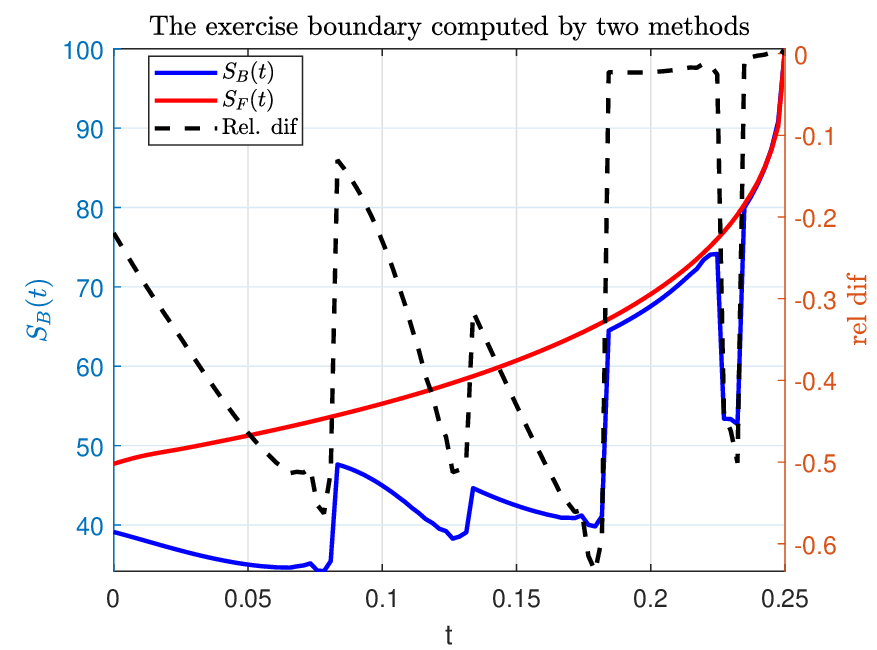}}
\end{center}
\caption{Early exercise boundaries for an American Put option under the GBM model, computed using our method with only proportional dividends: (a) time-dependent parameters, and (b) constant parameters (i.e., $r_1 = r_k = \sigma_1 = \sigma_k = 0$). The benchmark $S_F(t)$ is derived from a binomial tree assuming no discrete dividends and constant parameters, where the constant values are the time-average of the model's time-dependent parameters.}
\label{divPropSep}
\end{figure}

As predicted by classical theory, the early exercise boundary (EB) for a proportional dividend is effectively vertical, existing only immediately before the ex-dividend date and solely for deep in-the-money options. Fig.~\ref{divProp} demonstrates a similar (but slightly different) behavior in the presence of both continuous dividends and time-dependent model parameters. To isolate the specific effect of the proportional dividend, we present the EB for a discrete proportional dividend alone in Fig.~\ref{divPropSep}. The first plot shows results with time-dependent parameters, while the second uses constant parameters (i.e.,
$r_1 = r_k = \sigma_1 = \sigma_k = 0$). It can be seen that both continuous dividends and time-dependent parameters of the model could significantly affect a shape of the EB.

\subsection{The EB of an American Put with cash discrete dividends.}

This case is the most involved, as it entails solving a more complex nonlinear equation obtained by substituting the expression for $U'_x(\tau_{j},x)$  from \eqref{Uprime} into \eqref{VolEB2}.
\begin{align} \label{VolEB3}
- K e^{\tau + y(\tau)} & \left[ 1 + \erf\left( \frac{2\tau + y(\tau)}{2 \sqrt{\tau}}\right) \right] = \int_0^\tau \int_{y(s)}^\infty \frac{e^{-\frac{(\xi - y(\tau))^2}{4(\tau - s)}}}{2\sqrt{\pi(\tau - s)}} \frac{\xi - y(\tau)}{\tau-s}  \eta(s,\xi) e^{-s (\xi-y(s))^2} d\xi ds \\
&- \sum_{j=1}^{N_d} \bm{1}_{\tau_{j} < \tau} \frac{2 \alpha(\tau_{j})} {K \sigma^2(\tau_{j})} D_j \Lambda(\tau_{j}, y(\tau_{j})), \nonumber \\
\Lambda(\tau_{j}, y(\tau_{j})) &= \int_{y(\tau_{j})}^{\infty}
\frac{\xi - y(\tau)}{2\sqrt{\pi (\tau - \tau_{j})^3}} e^{-\frac{- (\xi-y(\tau))^2} {4(\tau-\tau_{j})}} \left[ - \zeta(\tau_{j}, \xi) e^{-\tau_{j}(\xi-y(\tau_{j}))^2}  + e^{-\xi} U'_\xi(\tau_{j},\xi) \right] d\xi.
\nonumber
\end{align}

Recall, that in the absence of discrete cash dividends, this equation reduces to the form we considered in earlier examples (i.e., all $D_j = 0$). Then, the algorithm of solving it with $D_j \ne 0$ is as follows:
\begin{myAlgorithm}{Solving \eqref{VolEB3}} \label{procDD}
    \item Define the ex-dividend dates $t_{j}$ and the corresponding cash dividend amounts $D_j$ for $j=1,\ldots,M$.
    \item Solve \eqref{VolEB3} as in the no-dividend case until the first ex-dividend date, $\tau \le \tau_{1-}$.
    \item At $\tau = \tau_{1-}+0$, compute the first term of the sum in \eqref{VolEB3}.
    \item Continue solving \eqref{VolEB3} with this new term included and updated at every $\tau$ until the next ex-dividend date, $\tau \le \tau_{2-}$.
    \item At $\tau = \tau_{2-}+0$, compute the second term of the sum in \eqref{VolEB3}.
    \item Repeat this process sequentially for all remaining dividend dates $j=3,\ldots,M$.
\end{myAlgorithm}

Thus, this differs from the case of no dividends by the necessity to compute $\Lambda(\tau_{j}, y(\tau_{j}))$ at times right after each ex-dividend date. Fortunately, a part of these calculations could be done analytically as this is shown in Appendix~\ref{appLambda}. The final result reads
\begin{align}
\Lambda(\tau_{j}, y(\tau_{j})) &= K \beta(\tau_{j}) \left[e^{-y(\tau_{j})} \left( a_0 J_2(1) + a_1 J'_2(1) + a_2 J''_2(1)\right) + b_0 J_2(0) + b_1 J'_2(0) + b_2 J''_2(0) \right],
\end{align}
where coefficients $a_i, b_i, i=0,1,2$  are defined in \eqref{J2-1} and the function $J_2(A)$ and its derivatives are given in \cref{J2A,intI1}. It is important that
$J_2(A)$ and its derivatives can be computed all together with just one evaluation of an exponential, a complementary error function ($\erfc$), and a square root.

In this test we maintain the same experimental setup as before, substituting discrete proportional dividends with discrete cash dividends. While the ex-dividend dates remain identical, the dividend amounts are now defined as fixed cash values: $[0.05, 0.04, 0.03, 0.02] \times K$. This formulation preserves the same nominal percentages as in the proportional case but links them to the strike price $K$ instead of the spot prices at the ex-dividend dates $S_{t-}$.
\begin{figure}[!htp]
\begin{center}
\hspace*{-0.3in}
\subfloat[]{\includegraphics[width=0.5\textwidth]{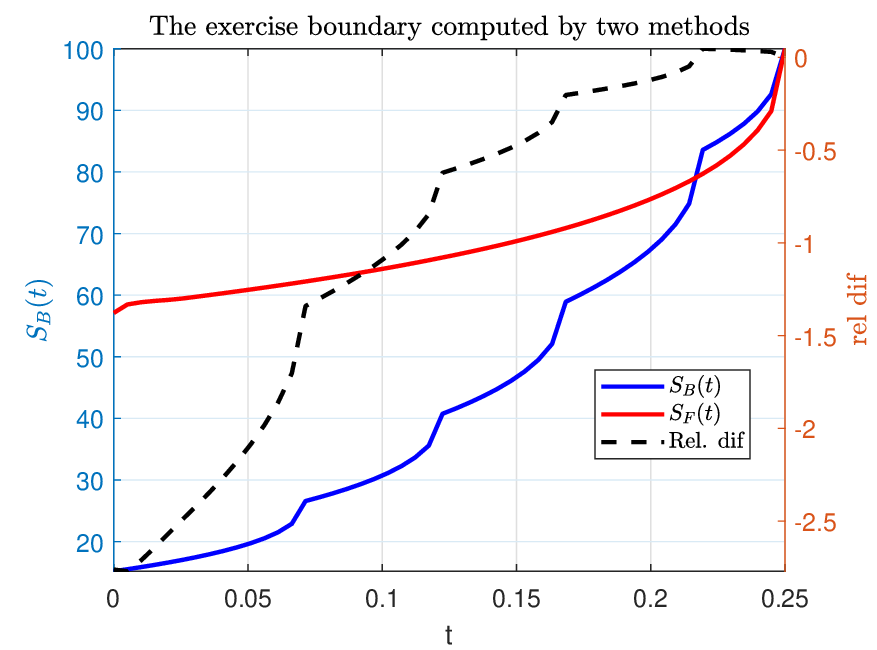}}
\subfloat[]{\includegraphics[width=0.5\textwidth]{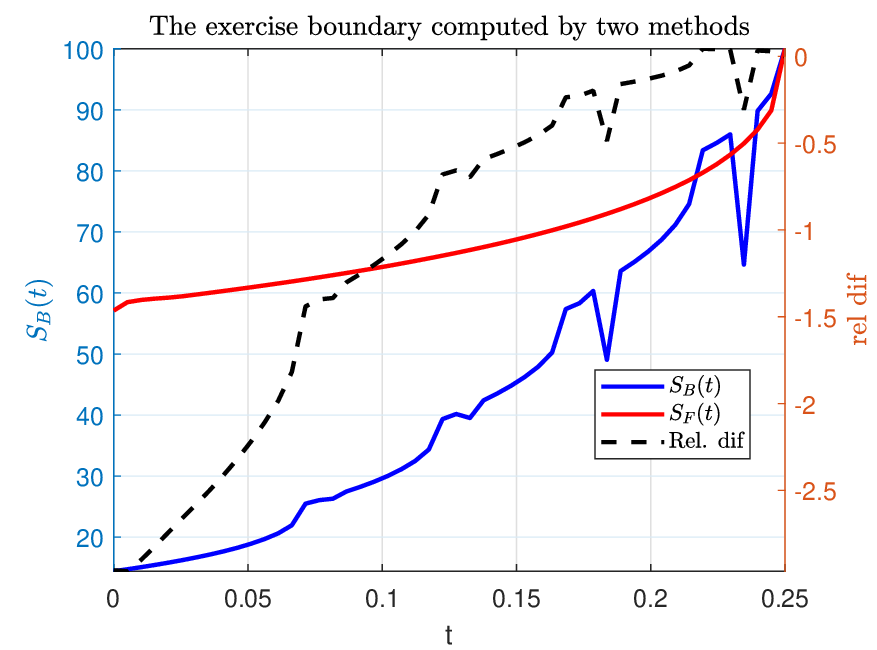}}
\end{center}
\caption{Early exercise boundaries $S_B(t)$ for an American put option under the time-inhomogeneous GBM model, computed using our method for two cases: (a) with continuous and discrete cash dividends, and (b) with all dividend types (continuous, discrete cash, and discrete proportional). The benchmark $S_F(t)$ is derived from a binomial tree assuming no discrete dividends and constant parameters, where the constant values are the time-average of the model's time-dependent parameters.}
\label{cashDiv}
\end{figure}

The results are presented in Fig.~\ref{cashDiv}(a). The typical elapsed time increases slightly to 0.085 seconds, as expected.

Finally, Fig.~\ref{cashDiv}(b) shows the early exercise boundary for an American Put where the underlying pays all types of dividends: continuous, discrete cash (as in the previous experiment), and discrete proportional dividends. The proportional dividends use the same amounts as in \cref{secPropDiv} but with slightly shifted ex-dividend dates (to distinguish them from the discrete cash dividends) to $[0.08, 0.13, 0.18, 0.23]$. The elapsed time again increases slightly, to 0.11 seconds.

It is worth noting that the current computation time is dominated by a non-vectorizable for-loop that calculates the EB using basic arithmetic operations (like MATLAB's \verb|fzero| solver. Due to the interpreter overhead in MATLAB loops, a C++ implementation would be dramatically faster. One could expect a speedup of 100x or more, which aligns with the performance reported in \cite{Andersen2016}.

\section{Conclusion} \label{conclusion}

This paper presents a novel, semi-analytical framework for pricing American options on assets that pay various types of dividends including discrete cash and proportional dividends (taken into account by using Dirac delta functions) and continuous dividends. The core challenge lies in the complex interplay between the optimal early exercise strategy and the predictable, discontinuous jumps in the underlying asset price at ex-dividend dates.

The proposed methodology is built upon two foundational pillars: a) the decomposition formula which represents the American option price as the sum of the corresponding European option price and the EEP, while the EEP is formulated as an integral that depends critically on the unknown EB; and b) the GIT method which, to determine the EB, transforms the problem into an non-linear Volterra integral equation of the first or second kind with respect to $S_B(t)$. This equation can be solved by various method including fixed-point iterations, etc.

To emphasize, the model accommodates a realistic market setting by incorporating both discrete cash and proportional dividends within a popular time-inhomogeneous Geometric Brownian Motion framework, featuring time-dependent interest rates, borrow costs, and volatility. A similar approach could be developed for some other one-factor models as this is shown in \cite{ItkinMuravey2024jd,ItkinCF2025}.

Key advantages of the proposed approach are:
\begin{itemize}
\item {\it Efficiency and accuracy}: By leveraging the structure of the integral equations and the FGT for numerical computation, the method achieves high accuracy with linear computational complexity in the spatial variable. This is a significant advantage over traditional methods like finite difference or binomial trees, which require dense temporal grids and lose efficiency, especially with multiple dividends.

\item {\it Handling singularities}: The paper provides a rigorous treatment of the numerical challenges posed by the problem, including the weak singularities inherent in the Volterra equations and the known singularity in the derivative of the EEB at expiration. A change of variables is proposed to effectively remove these singularities, ensuring robust numerical solutions.

\item {\it Simultaneous solution for strikes and maturities}: A particularly powerful insight is the method's potential for "de-Americanization" – converting American option prices into their European equivalents for use in local volatility models. By using a notion of "impled strike", \cite{Skabelin2015}, the structure of the equations allows for the simultaneous computation of implied strikes across multiple maturities and market quotes in a single computational pass, a task that is computationally prohibitive for traditional iterative methods.
\end{itemize}

The analysis also yields important theoretical insights. The derived integral equations for the EEB formally demonstrate the well-known financial results, namely: for American Put options, the early exercise boundary ceases to exist (i.e., early exercise is never optimal) when interest rates are negative. Conversely, for American call options, the early exercise boundary does not exist when interest rates are positive. This is derived mathematically from the properties of the governing integral equations.

In summary, this work provides a robust, efficient, and theoretically sound semi-analytical alternative to purely numerical methods for a classic problem in quantitative finance. By reformulating the problem in terms of integral equations and employing advanced analytics and numerical techniques (like the FGT for computing integrals with Gaussian kernels), it offers a path to accurate pricing and risk management of American options with discrete dividends, even in demanding applications like calibration of local volatility surfaces from market data. The framework is general and can be extended to more complex underlying dynamics beyond the time-inhomogeneous GBM model considered here.

\section*{Acknowledgments}

\noindent I thank Leif Andersen for fruitful discussions and Alex Skabelin for useful comments. I am also grateful to an anonymous referee for their helpful feedback. 

\vspace{\baselineskip} 
\noindent The use of LLMs in this paper has been limited to minimal proofreading. Both the mathematical content and the drafting of this research have been produced without any kind of reliance on generative AI.

\section*{Conflict of interest}

The author has no competing interests to declare that are relevant to the content of this article.

\section*{Funding}

The author has no relevant financial or non-financial interests to disclose.

\section*{Compliance with Ethical Standards}

This article does not contain any studies with human participants or animals performed by any of the authors.

\vspace{0.4in}
\printbibliography[title={References}]

\vspace{0.4in}
\appendixpage
\appendix
\numberwithin{equation}{section}
\setcounter{equation}{0}

\section{Computing the transition density step-by-step} \label{app1}

This case differs from the European put option pricing problem described in \cref{Europ}, namely: while we still need to solve the PDE in \eqref{PDE2}, the source term $\Phi(\tau,x)$ now vanishes, and the initial condition in \eqref{tcP} becomes a delta function. Accordingly, instead of \eqref{Volt2} we now obtain the following Volterra-Fredholm equation of the second kind
\begin{align} \label{Volt2N}
W(\tau, x|0,x_0) &= \frac{e^{-\frac{(x-x_0)^2}{4 \tau }}}{2 \sqrt{\pi \tau }} \\
&+ \sum_{j=1}^{N_d} D_j \bm{1}_{\tau_{j} \le \tau} \int_{-\infty}^{\infty} \frac{x - \xi - 2 (\tau - \tau_{j})}{\tau - \tau_{j}} \frac{\alpha(\tau_{j})}{\sigma^2(\tau_{j})} e^{-\xi} G(x, \xi, \tau-\tau_{j}) W(\tau_{j},\xi|0,x_0) d \xi, \nonumber
\end{align}
where the Green's function $G(x, \xi, \tau)$ remains defined by \eqref{Green}. Here, discrete cash dividends are counted forward in time $\tau$. The notation $\bm{1}_{\tau_{j} \le \tau}$ indicates that we only consider dividends satisfying $0 < \tau_{1-} < \ldots < \tau_{k-} < \tau$, where $\tau_j$ is the inverse map of $T_j$ as defined in \eqref{tauP}.

To solve \eqref{Volt2N} we proceed as follows:

\begin{myAlgorithm}{Solving \eqref{Volt2N}} \label{procP}
\item $\bm{\tau < \tau_{1-}}$. The solution of \eqref{Volt2N} coincides with the log-normal density same as in the Black-Scholes model with time-dependent coefficients (the first term in the right-hands part of \eqref{Volt2N}).

\item $\bm{\tau = \tau_{1-}}$. Using the identity \eqref{deltaPrime}, we obtain from \eqref{Volt2N}
\begin{align} \label{unmodAP}
W(\tau_{1-}, x|0,x_0) &= A(\tau_{1-}, x) + B(\tau_{1-}) e^{-x} \left[ W'_x(\tau_{1-},x|0,x_0) - 2 W(\tau_{1-},x|0,x_0)\right], \\
A(\tau,x) &= \frac{e^{-\frac{(x-x_0)^2}{4 \tau }}}{2 \sqrt{\pi \tau }}. \nonumber
\end{align}
This is a first-order ordinary differential equation for $W(\tau_{1-},x|0,x_0)$. By solving it, we obtain
\begin{equation} \label{Wdiv1}
W(\tau_{1-},x|0,x_0) = e^{\frac{e^x}{B(\tau_{1-})} + 2 x} \frac{1}{B(\tau_{1-})}\int_{x}^\infty A(\tau_{1-},k) e^{-k - e^{k}/B(\tau_{1-})} dk.
\end{equation}
At $x \to \pm \infty$ this yields $W(\tau_{1-},x|0,x_0) \to 0$ which are the correct boundary conditions, see \eqref{bcP}.

\item $\bm{\tau_{2-} > \tau > \tau_{1-}}$. Since $W(\tau_{1-},x|0,x_0)$ has already been found at the previous step, the transition density $W(\tau,x|0,x_0)$ is again given by the log-normal density identical to that in the Black-Scholes model with time-dependent coefficients (the first term on the right-hands part of \eqref{Volt2N}), plus the second integral where all integrand terms are known from prior calculations.

\item $\bm{\tau = \tau_{2-}}$. This step is similar to the step (b) with the only difference that $A(\tau_{2-},x)$ is now defined as
\begin{align} \label{Adiv1}
A (\tau_{2-},x) &= \frac{e^{-\frac{(x-x_0)^2}{4 \tau }}}{2 \sqrt{\pi \tau }} + \frac{1}{2\sqrt{\pi}} \sum_{j=1}^{1} D_j \calJ_j(\tau_{2-},x), \\
\calJ_j(\tau_{2-},x) &= \int_{-\infty}^{\infty} \frac{x - \xi + 2 (\tau_{2-} - \tau_{1-})}{\tau_{2-} - \tau_{1-}} \frac{\alpha(\tau_{2-})}{\sigma^2(\tau_{2-})} e^{-\xi} G(x, \xi, \tau_{2-}-\tau_{1-}) W(\tau_{1-},\xi|0,x_0) d \xi. \nonumber 
\end{align}

\item and so on ...
\end{myAlgorithm}

Again, the integrals in \cref{Wdiv1,Adiv1} can be computed with linear complexity by using \cite{FGT2010,Greengard2024}.  Alternatively, similar to the computation of the integral in \eqref{I2}, we can rewrite \eqref{Wdiv1} as
\begin{align} \label{TaylorP}
W & (\tau,x|0,x_0) = \frac{e^{2x}}{B(\tau)} \int_{x}^{\infty} A(\tau,k) e^{-k}\left[ e^{-e^{x - L_B}} - e^{-e^{k - L_B}} \right] = e^{-e^{-L_B + x} - L_B + x} \frac{e^{2x}}{B(\tau)}  \sum_{i=1}^N a_i(L_B,x) \\
&\cdot \int_x^\infty A(\tau,k) (k - x)^i e^{-k} dk = \frac{e^{2x}}{B(\tau)} \frac{2^{k-1} \tau^{k/2}}{\sqrt{\pi}}
e^{- x - \frac{(x-x_0)^2}{4 \tau}} \Bigg[ \Gamma \left(\frac{k+1}{2}\right) \, _1F_1\left(\frac{k+1}{2};\frac{1}{2};\frac{(x-x_0+2 \tau)^2}{4 \tau}\right) \nonumber \\
&+ \frac{x-x_0+2 \tau}{\sqrt{\tau}} \Gamma \left(\frac{k}{2}+1\right)\, _1F_1\left(\frac{k+2}{2};\frac{3}{2};\frac{(x-x_0+2 \tau)^2}{4 \tau}\right) \Bigg], \qquad L_B = \log(B(\tau)), \quad \tau = \tau_{1-}, \nonumber
\end{align}
where $\Gamma(x)$ is the gamma function, $_1F_1(a,b,x)$ is the Kummer confluent hypergeometric function, \cite{as64}.

In \eqref{TaylorP} we employ a Taylor series expansion of $e^{-e^{x - L_B}} - e^{-e^{k - L_B}}$ about $k = x$, retaining terms up to order $N$. The coefficients of this expansion can be readily computed, e.g.
\begin{equation}
a_1(L_B,x) = 1, \quad a_2(L_B,x) = \frac{1}{2} \left(e^{L_B} - e^x\right) e^{-L_B}, ...
\end{equation}
However, computing the integral in \eqref{Adiv1} and subsequently evaluating the expectations in \eqref{decompGen2} may be computationally expensive when using analytical representations of the integrands in terms of special functions. Therefore, numerical integration via the FGT method is preferable.

\section{Integral equation for the exercise boundary of the American Put} \label{appPsi}

To derive an integral Volterra equation of the first kind for the EB, similar to \cite{ItkinMuravey2024jd} we first, differentiate both parts of \eqref{solPut} on $x$, and then set $x \to y(\tau)$. The first step yields
\begin{align} \label{solPutX}
\Psi(\tau, x) &= \frac{1}{2\sqrt{\pi \tau}} \int_{y(0)}^{\infty} U(0,\xi) \left[ -\frac{y(\tau) - \xi}{2\tau} e^{-\frac{(y(\tau) - \xi)^2}{4\tau}} + \frac{\xi - y(\tau)}{2\tau} e^{-\frac{(y(\tau) - \xi)^2}{4\tau}} \right] \\
&+ \int_{0}^{\tau} \frac{\Psi(s, y(s))}{4\sqrt{\pi(\tau - s)^3}} \left[ (x-y(s)) e^{-\frac{(x - y(s))^2}{4(\tau - s)}} -  (x + y(s) - 2y(\tau))) e^{-\frac{(x - 2y(\tau) + y(s))^2}{4(\tau - s)}} \right] ds \\
&- \int_{0}^{\tau} \int_{y(s)}^{\infty} \frac{\lambda(s, \xi)}{4\sqrt{\pi(\tau - s)^3}} \left[ (x-\xi) e^{-\frac{(x-\xi)^2}{4(\tau - s)}} - (x + \xi - 2y(\tau)) e^{-\frac{(\xi - 2y(\tau) + x)^2}{4(\tau - s)}} \right] d\xi ds. \nonumber
\end{align}
Here, the function $\Psi(\tau, x)$ is the gradient of the solution
\begin{equation}
\Psi(\tau, x) =  \fp{U(\tau, x)}{x}.
\end{equation}
We now need to take the limit $x \to y(\tau)$. As detailed in \cite{TS1963,ItkinLiptonMuraveyBook}, for the first two integrals the result depends on whether we approach the EB from the right, $x \to y(\tau)+0$, or from the left, $x \to y(\tau)-0$. But, since we are working within the continuation region of the American Put, which is $x \in [y(\tau), \infty)$, we require the limit from the right: $x \to y(\tau)+0$.

We begin with the first part of the first integral $I_1$ in \eqref{solPutX}. Using the result in \cite{TS1963,ItkinLiptonMuraveyBook}, we obtain
\begin{align}
\lim_{x \to y(\tau)+0} \int_{0}^{\tau} \frac{\Omega(s, y(s))}{4\sqrt{\pi(\tau - s)^3}} (x-y(s)) e^{-\frac{(x - y(s))^2}{4(\tau - s)}} ds &= \frac{1}{2} \Omega(\tau, y(\tau)) + \int_{0}^{\tau} \frac{\Omega(s, y(s))}{4\sqrt{\pi(\tau - s)^3}} (y(\tau)-y(s)) e^{-\frac{(y(\tau) - y(s))^2}{4(\tau - s)}}, \nonumber \\
\Omega(s, y(s)) &\equiv \Psi(s, y(s)).
\end{align}

Similarly, for the second part of the first integral in \eqref{solPutX}, introducing $\bar{x} = 2 y(\tau) - x$, we have
\begin{align}
-\lim_{x \to y(\tau)+0} & \int_{0}^{\tau} \frac{\Omega(s, y(s))}{4\sqrt{\pi(\tau - s)^3}} (2 y(\tau) - x - y(s)) e^{-\frac{(2 y(\tau) - x - y(s))^2}{4(\tau - s)}} ds \\
&= -\lim_{\bar{x} \to y(\tau)-0} \int_{0}^{\tau} \frac{\Omega(s, y(s))}{4\sqrt{\pi(\tau - s)^3}} (\bar{x} - y(s)) e^{-\frac{(\bar{x} - y(s))^2}{4(\tau - s)}} ds \nonumber \\
&=  -\frac{1}{2} \Omega(\tau, y(\tau)) + \int_{0}^{\tau} \frac{\Omega(s, y(s))}{4\sqrt{\pi(\tau - s)^3}} (y(\tau)-y(s)) e^{-\frac{(y(\tau) - y(s))^2}{4(\tau - s)}}. \nonumber
\end{align}
Thus,
\begin{align}
\lim_{x \to y(\tau)+0} I_1 = \int_{0}^{\tau} \frac{\Omega(s, y(s))}{2\sqrt{\pi(\tau - s)}} \frac{y(\tau)-y(s)}{\tau - s} e^{-\frac{(y(\tau) - y(s))^2}{4(\tau - s)}} ds.
\end{align}

%

This result can also be directly applied to the second integral in \eqref{solPutX} when $\xi = y(s)$ while for other values of $\xi$ there is no such an issue.

For the first integral in \eqref{solPutX} we have
\begin{align}
J_0 &= \int_{y(0)}^\infty U(0,\xi) \frac{\xi - y(\tau)}{2\sqrt{\pi \tau ^3}} e^{-\frac{(\xi - y(\tau))^2}{4\tau}} d\xi  = -
\int_{y(0)}^\infty \frac{ U(0,\xi) }{\sqrt{\pi \tau}} d \left(e^{-\frac{(\xi - y(\tau))^2}{4\tau}} \right) \\
&= \frac{U(0,y(0))}{\sqrt{\pi \tau}} e^{-\frac{(y(0) - y(\tau))^2}{4\tau}} + \int_{y(0)}^\infty \frac{ U'_\xi(0,\xi)}{\sqrt{\pi \tau}} e^{-\frac{(\xi - y(\tau))^2}{4\tau}} d\xi = \frac{1}{\sqrt{\pi \tau}}\int_{y(0)}^\infty U'_\xi(0,\xi) e^{-\frac{(\xi - y(\tau))^2}{4\tau}} d\xi, \nonumber
\end{align}
since $U(0,y(0)) = 0$.


Further, using the definition of $h(\tau,x)$ in \eqref{PDE_EB}, an expression for $\Psi(\tau, y(\tau)$ in \eqref{sub} and combining them with the above results together, we finally obtain a nonlinear integral Volterra equation of the first kind for $y(\tau)$
\begin{align} \label{psiFin}
0 &= \frac{1}{\sqrt{\pi \tau}}\int_{y(0)}^\infty U'_\xi(0,\xi) e^{-\frac{(\xi - y(\tau))^2}{4\tau}} d\xi +  \int_0^\tau \int_{y(s)}^\infty \frac{e^{-\frac{(\xi - y(\tau))^2}{4(\tau - s)}}}{2\sqrt{\pi(\tau - s)}} \frac{\xi - y(\tau)}{\tau-s}  \eta(s,\xi) e^{-s(\xi-y(s))^2} d\xi ds, \\
&- \int_0^\tau \gamma(s) \int_{y(s)}^{\infty} \left[ e^{-\xi} U'_\xi(s,\xi) - \zeta(s,\xi) e^{-s (\xi-y(s))^2} \right] \frac{\xi - y(\tau)}{2\sqrt{\pi (\tau - s)^3}} e^{-\frac{(\xi - y(\tau))^2}{4(\tau-s)}} d\xi ds, \nonumber
\end{align}

\section{Derivation of $\calJ_3(\tau_{j},\tau_{j},x)$ in \eqref{solPut2}} \label{appTj}

Recall that by the definition in \eqref{solPut2}
\begin{align} \label{solPut2-1}
\calJ_3(\tau_{j}, \tau_{i},x) &=  \int_{y(\tau_{i})}^{\infty} \zeta(\tau_{i},\xi) \tjm14 e^{-\frac{(\xi-y(\tau_{i}))^2}{\sqrt{\tau_{i}}}} \frac{\xi - y(\tau_{j})}{2\sqrt{\pi (\tau_{j} - \tau_{i})^3}} e^{-\frac{(\xi - y(\tau_{j}))^2}{4(\tau_{j}-\tau_{i})}} d\xi \\
&+ \int_{y(\tau_{i})}^{\infty} e^{-\xi} U(\tau_{i},\xi) \left\{ - \frac{1}{2\sqrt{\pi(\tau_{j} - \tau_{i})}} \left[ e^{-\frac{(x-\xi )^2}{4 (\tau_{j}-\tau_{i})}} - e^{-\frac{(\xi +x-2 y(\tau_{j} ) )^2}{4 (\tau_{j}-\tau_{i})}} \right] \right. \nonumber \\
&+ \left. \frac{1}{4\sqrt{\pi(\tau_{j} - \tau_{i})^3}} \left[(x-\xi)e^{-\frac{(x-\xi )^2}{4 (\tau_{j}-\tau_{i})}} + (x + \xi - 2y(\tau_{j})) e^{-\frac{(\xi +x-2 y(\tau_{j} ))^2}{4 (\tau_{j}-\tau_{i})}}  \right] \right\} d\xi, \qquad i < j. \nonumber
\end{align}
Further, for ease of notation we drop the index in $\tau_{j}$. At $i=j$ the second integral takes the form
\begin{align} \label{solPut2-2}
\calJ_3(\tau, \tau,x) &= \int_{y(\tau)}^{\infty} e^{-\xi} U(\tau,\xi) \left[ \delta(\xi +x-2 y(\tau)) - \delta(x - \xi) + \delta'_\xi(x-\xi) - \delta'_\xi(\xi +x-2 y(\tau)) \right].
\end{align}
Thus, the result is a sum of four integrals. Consider, for instance, the last one
\begin{equation}
I_4 = \int_{y(\tau )}^{\infty } f(\xi) \frac{\partial \delta (x+\xi -2 y(\tau ))}{\partial \xi } \, d\xi, \qquad f(\xi) = -e^{-\xi} U(\tau, \xi).
\end{equation}

To simplify the argument of the delta function, let's perform a change of variables: $u = x+\xi - 2y(\tau)$. Substituting these into the integral, we get:
\begin{equation}
I_4 = \int_{x-y(\tau )}^{\infty } f(u - x + 2y(\tau)) \delta'(u) \, du
\end{equation}
By the sifting property of the derivative of the Dirac delta function
\begin{equation}
\int_a^b g(u) \delta'(u) \, du = -g'(0).
\end{equation}
This identity holds if and only if the point $u=0$ is within the interval of integration $[a, b]$. Otherwise, if $u=0$ is outside this interval, the integral is zero. Since in our case $x > y(\tau)$, we have $I_4 = 0$.

In a similar manner, we get
\begin{align}
I_3 &= \int_{y(\tau )}^{\infty } f(\xi) \frac{\partial \delta(x - \xi)}{\partial \xi } \, d\xi = -f'(x), \qquad
I_2 = \int_{y(\tau )}^{\infty } f(\xi) \delta(x + \xi - 2 y(\tau)) \, d\xi = 0, \\
I_1 &= -\int_{y(\tau )}^{\infty } f(\xi) \delta(x - \xi) \, d\xi = -f(x). \nonumber
\end{align}

The first integral can be calculated in a similar manner since at $i = j$ the integrand is also proportional to the derivative of a delta function. Collecting all terms, this yields
\begin{equation}
\calJ_3(\tau_{j}, \tau_{j},x) = - \zeta_\xi(\tau_{j}, y(\tau_{j})) - e^{-x} U'(\tau_{j},x).
\end{equation}

\section{Calculation of the internal integral in \eqref{VolEB-a1}} \label{appGA}

Here, we want to calculate the integral
\begin{align}
I = \int_{y(s)}^\infty \frac{e^{-\frac{(\xi - y(\tau))^2}{4(\tau - s)}}}{2\sqrt{\pi(\tau - s)}} \frac{\xi - y(\tau)}{\tau-s}  \eta(s,\xi) e^{-s (\xi-y(s))^2} d\xi,
\end{align}
where $\eta(\tau,x)$ is defined in \eqref{chgU} as
\begin{align} \label{etaZ}
\eta(s,\xi) &= z(s,\xi) \left[ Z^2 (1 + 4 s^2)  - 2 s y'(s) Z - (2 s + \bar{r}'(s)) \right] + K \beta(s) e^\xi \left[4 s Z -  \rho'(s)\right], \quad Z = \xi - y(s).
\end{align}

Using the representation
\begin{align}
 \frac{\xi - y(\tau)}{2(\tau-s)} & e^{-\frac{(\xi - y(\tau))^2}{4(\tau - s)}} e^{- s (\xi-y(s))^2}
= - \calL \left[ e^{-\frac{(\xi - y(\tau))^2}{4(\tau - s)} - s (\xi-y(s))^2} \right], \qquad \calL = \fp{}{\xi}
+ 2 s (\xi - y(s)),
\end{align}
and integrating by parts, the double integral in \eqref{VolEB-a1} can be represented as
\begin{align}
I &= - \int_0^\tau \int_{y(s)}^\infty \frac{\eta(s,\xi)}{\sqrt{\pi(\tau - s)}} \calL \left[ e^{-\frac{(\xi - y(\tau))^2}{4(\tau - s)} - s (\xi-y(s))^2} \right] d\xi ds = -(I_1 + I_2), \\
I_2 &= 2 \int_0^\tau \int_{y(s)}^\infty \frac{s \eta(s,\xi)(\xi - y(s))}{\sqrt{\pi(\tau-s)}}  e^{-\frac{(\xi - y(\tau))^2}{4(\tau - s)} - s(\xi-y(s))^2}, \nonumber\\
I_1 &= - \int_0^\tau \eta(s,y(s)) \frac{e^{-\frac{(y(s) - y(\tau))^2}{4(\tau - s)}}}{\sqrt{\pi(\tau - s)}}  ds -
\int_0^\tau \int_{y(s)}^\infty \frac{\eta'_{\xi}(s,\xi)}{\sqrt{\pi(\tau - s)}} e^{-\frac{(\xi - y(\tau))^2}{4(\tau - s)} - s(\xi-y(s))^2} d\xi ds. \nonumber
\end{align}

Combining all these expressions together, yields
\begin{align} \label{intI}
I &= \int_0^\tau \eta(s,y(s)) \frac{e^{-\frac{(y(s) - y(\tau))^2}{4(\tau - s)}}}{\sqrt{\pi(\tau - s)}} ds + \int_0^\tau \int_{y(s)}^\infty \frac{\eta'_{\xi}(s,\xi) - 2 s \eta(s,\xi)(\xi-y(s))}{\sqrt{\pi(\tau - s)}} e^{-\frac{(\xi - y(\tau))^2}{4(\tau - s)} - s(\xi-y(s))^2} d\xi ds.
\end{align}

Therefore, now we need to calculate
\begin{equation*}
J_1 = \int_{y(s)}^\infty \frac{\eta'_{\xi}(s,\xi) - 2 s \eta(s,\xi)(\xi-y(s))}{\sqrt{\pi(\tau - s)}} e^{-\frac{(\xi - y(\tau))^2}{4(\tau - s)} - s(\xi-y(s))^2} d\xi = \frac{e^{-\frac{(y(s) - y(\tau))^2}{4(\tau - s)}}}{\sqrt{\pi(\tau - s)}} \int_{0}^\infty F(s,Z) e^{- \frac{p_2 Z^2 + p_1 Z}{4(\tau - s)}} dZ,
\end{equation*}
\vspace{-1em}
\begin{equation} \label{J1}
F(s,Z) = \eta'_{Z}(s,Z) - 2 s Z \eta(s,Z) = K \beta(s) \sum_{i=0}^3 \left[a_i e^{Z + y(s)} + b_i\right] Z^i,  \\
\end{equation}
\vspace{-1em}
\begin{alignat}{2}
a_0 &= \bar{r}'(s)  - \rho'(s) +2 s \left(y'(s)+3\right), &\qquad
a_1 &= 2 \left(s \left(2 - \bar{r}'(s) + \rho'(s) + y'(s)\right) - 6 s^2 - 1\right), \nonumber \\
a_2 &= - (1 + 4 s^2 (3 + y'(s))), &\qquad a_3 &= 2 s \left(4 s^2+1\right), \nonumber \\
b_0 &= -2 s \alpha(s) y'(s), &\qquad b_1 &= 2 \alpha(s)\left(1 + 6 s^2 + s \bar{r}'(s) \right), \nonumber \\
b_2 &= 4 s^2 \alpha(s) y'(s), &\qquad b_3 &= -2 s\alpha(s) \left(4 s^2+1\right). \nonumber
\end{alignat}

In turn, using a standard trick, the last integral can be re-written in the form
\begin{align} \label{J10}
J_{1,0} &= \int_{0}^\infty F(s,Z) e^{- \frac{p_2 Z^2 + p_1 Z}{4(\tau - s)}} dZ = K \beta(s) \left[e^{y(s)} \lim_{A \to 1} \sum_{i=0}^3 a_i + \lim_{A \to 0} \sum_{i=0}^3 b_i \right] \frac{\partial^i}{\partial A^i}  \int_{0}^\infty e^{- k_2 Z^2 + k_1(A) Z} dZ, \nonumber \\
k_1(A) &= A + \frac{y(\tau) - y(s)}{2(\tau-s)}, \qquad k_2 =  s + \frac{1}{4(\tau-s)}.
\end{align}

The integral in \eqref{J10} can be computed in closed form to yield
\begin{equation} \label{J2A}
J_2(A) = \int_{0}^\infty e^{- k_2 Z^2 + k_1(A) Z} dZ = \frac{\sqrt{\pi}}{2 \sqrt{k_2}} e^{\frac{k_1(A)^2}{4 k_2}} \left[1 + \erf\left(\frac{k_1(A)}{2 \sqrt{k_2}}\right)\right].
\end{equation}
Accordingly,
\begin{align} \label{intI1}
J'_2(A) &= \frac{1}{2 k_2} + \frac{\sqrt{\pi} k_1(A)}{4 k^{3/2}_2} e^{\frac{k_1^2(A)}{4 k_2}} \left[1 + \erf\left(\frac{k_1(A)}{2 \sqrt{k_2}} \right) \right], \\
J''_2(A) &= \frac{k_1(A)}{4 k^2_2} + \frac{\sqrt{\pi} (2 k_2 + k_1^2(A))}{8 k^{5/2}_2} e^{\frac{k_1^2(A)}{4 k_2}} \left[1 + \erf\left(\frac{k_1(A)}{2 \sqrt{k_2}} \right) \right], \nonumber \\
J'''_2(A) &= \frac{4k_2 +  k_1^2(A)}{8 k^3_2} + \frac{\sqrt{\pi} k_1(A)(6 k_2 + k_1^2(A))}{16 k^{7/2}_2} e^{\frac{k_1^2(A)}{4 k_2}} \left[1 + \erf\left(\frac{k_1(A)}{2 \sqrt{k_2}} \right) \right], \nonumber
\end{align}
and, hence
\begin{align}
J_{1,0} &= K \beta(s) \left[e^{y(s)} \left( a_0 J_2(1) + a_1 J'_2(1) + a_2 J''_2(1) + a_3 J'''_2(1) \right) + b_0 J_2(0) + b_1 J'_2(0) + b_2 J''_2(0) + b_3 J'''_2(0) \right].
\end{align}

Finally, combining \cref{intI,intI1}, we obtain
\begin{align} \label{intIFin}
I &= \int_0^\tau \frac{e^{-\frac{(y(s) - y(\tau))^2}{4(\tau - s)}}}{\sqrt{\pi(\tau - s)}} \left[ \eta(s,y(s)) + J_{1,0} \right] ds,
\end{align}
where
\begin{equation}
\eta(s,y(s)) = - (2s + \barr'(s)) z(s,y(s)) - K \beta(s) e^{y(s)} \rho'(s).
\end{equation}

One can verify that the dominant contribution to the integrand in \eqref{intIFin} comes from the first term inside the brackets. Indeed, consider both terms in the limit $s \to \tau$. By the definition of $J_2$ and its derivatives in  \eqref{intI1}, at $s \to \tau$ $J_2$ vanishes together with all its derivatives, while the first term in the braces $\eta(\tau, y(\tau)$ not.
\begin{figure}[!htp]
\centering
\fbox{\includegraphics[width=0.7\textwidth]{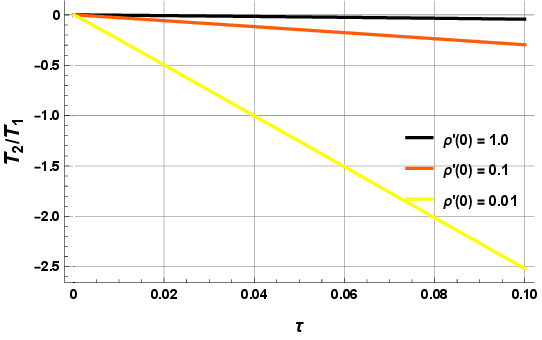}}
\caption{Comparison of two terms $T_1$ and $T_2$ in braces in \eqref{intIFin} at various values of $\rho'(0)$.}
\label{dominance}
\end{figure}

Similar analysis can be done as $s \to 0$. In this limit we have $a_0 = a_3 = b_0 = b_2 = b_3 = 0$, $b_1 = 2, a_1 = -2, a_2 = -1$,and all non-vanished terms in $J_{1,0}$ are of the order of
\begin{equation*}
T_2 = \frac{1}{16}K \sqrt{\pi \tau} e^{\frac{y(\tau )^2}{64 \tau}} y(\tau ) \left[1 + \erf\left(\frac{y(\tau)}{8\sqrt{\tau }}\right) \right],
\end{equation*}
while the first term in brackets in \eqref{intIFin} reads
\begin{equation*}
T_1 = - K \rho'(0).
\end{equation*}

The dominance of the $T_1$ term, which is the motivation for isolating it, occurs when $\rho'(0) \sim O(1)$ and so $T_1 \gg T_2$, a relationship confirmed numerically in Fig.~\ref{dominance}. Here, we use an asymptotic behavior of $y(\tau)$ as in \eqref{asympt}
\begin{equation}
y(\tau) \approx -\tau + \rho'(\tau) \tau - \sqrt{-\tau \log \left(2 \pi  \rho'(\tau)^2 \tau \right)},
\end{equation}
and compute the ratio $T_2/T_1$ as a function of $\tau \in [0,0.1]$ (the upper limit is chosen to be able to apply the asymptotic in \eqref{asympt}).

However, this hierarchy reverses for small $\rho'(0) \sim O(\tau)$ causing the second term to dominate. Such small values of $\rho'(s)$ are typically observed in environments characterized by high volatility and low instantaneous interest rates.

\section{Calculation of $\Lambda(\tau_{j}, y(\tau_{j}))$ in \eqref{VolEB3}} \label{appLambda}

As established previously, our focus is on the integral
\begin{align} \label{VolEB3-1}
\Lambda(\tau_{j}, y(\tau_{j})) &= \int_{y(\tau_{j})}^{\infty}
\frac{\xi - y(\tau)}{2\sqrt{\pi (\tau - \tau_{j})^3}} e^{-\frac{(\xi-y(\tau))^2} {4(\tau-\tau_{j})}} \left[ e^{-\xi} U'_\xi(\tau_{j},\xi) - \zeta(\tau_{j}, \xi) e^{-\tau_{j}(\xi-y(\tau_{j}))^2} \right] d\xi,
\end{align}
where, based on the definitions in \eqref{chgU}
\begin{align}
\zeta(\tau,x) &= 2 \tau z(\tau,x) (x - y(\tau))e^{-x} + K \beta(\tau), \qquad
\zeta'_x(\tau,x) = 2 \tau e^{-x} \left[z(\tau,x) - (x-y(\tau)) \alpha(\tau)\right].
\end{align}

Thus, $\Lambda(\tau_{j}, y(\tau_{j}))$ is a difference of two integrals, i.e. $\calJ_1 - \calJ_2$.

\subsection{Computation of $\calJ_2$}

This integral corresponds to the second term in square brackets in \eqref{VolEB3-1} and has the same integral structure as in \cref{appGA}. Therefore, it could be computed using the same approach. This yields
\begin{align} \label{J2def}
\calJ_2 &= \int_0^\infty \frac{e^{-\frac{(Z + y(\tau_{j}) - y(\tau))^2}{4(\tau - \tau_{j})} -\tau_{j} Z^2}}{\sqrt{\pi(\tau - \tau_{j})}} \left[ \zeta(\tau_{j},Z+y(\tau_{j})) + F(\tau_{j},Z + y(\tau_{j}))\right] dZ, \\
\zeta(\tau_{j},x) &+ F(\tau_{j},x)\Big|_{x \to Z + y(\tau_{j})} = \zeta'_{x}(\tau_{j},x) + 2 \left(1 - \tau_{j} x \right) \zeta(\tau_{j},x)\Big|_{x \to Z + y(\tau_{j})} \nonumber \\
&= K \beta(s) \sum_{i=0}^2 \left[e^{-y(\tau_{j})} a_i e^{-Z} + b_i\right] Z^i, \nonumber \\
a_0 &= 2 \tau_{j} \alpha(\tau_{j}), \qquad a_1 = 2 \tau_{j}\alpha(\tau_{j}) \left[1-2 \tau_{j} y(\tau_{j})\right], \qquad a_2 = - 8 \tau^2_{j} \alpha(\tau_{j}), \nonumber \\
b_0 &= -2 \left[ -1 + \tau_{j}(1 + y(\tau_{j})) \right], \qquad b_1 = - 2 \tau_{j} \left[ 3 - 2 \tau_{j} y(\tau_{j}) \right], \qquad b_2 = 8 \tau^2_{j}. \nonumber
\end{align}

In particular, by representing it in the form of $J_{1,0}$ in \eqref{J10}, we obtain
\begin{align} \label{J2-1}
\calJ_2 &= -K \beta(\tau_{j}) \frac{e^{-\frac{(y(\tau_{j}) - y(\tau))^2}{4(\tau - \tau_{j})}}}{\sqrt{\pi(\tau - \tau_{j})}}  \left[e^{-y(\tau_{j})} \lim_{A \to 1} \sum_{i=0}^1 (-1)^i a_i + \lim_{A \to 0} \sum_{i=0}^1 b_i \right] \frac{\partial^i}{\partial A^i}  \int_{0}^\infty e^{- k_2 Z^2 + k_1(A) Z} dZ, \nonumber \\
k_1(A) &= -A + \frac{y(\tau) - y(\tau_{j})}{2(\tau-\tau_{j})}, \qquad k_2 =  \tau_{j} + \frac{1}{4(\tau-\tau_{j})}.
\end{align}

Using \cref{J2A,intI1} gives the final result
\begin{align}
\calJ_2 &= K \beta(\tau_{j}) \left[e^{-y(\tau_{j})} \left( a_0 J_2(1) + a_1 J'_2(1) + a_2 J''_2(1)\right) + b_0 J_2(0) + b_1 J'_2(0) + b_2 J''_2(0) \right].
\end{align}

\subsection{Computation of $\calJ_1$}

The first integral in \eqref{VolEB3-1} can be simplified using integration by parts
\begin{align} \label{J2Z}
\calJ_1 &= \int_{y(\tau_{j})}^{\infty} \frac{\xi - y(\tau)}{2\sqrt{\pi (\tau - \tau_{j})^3}} e^{-\frac{(\xi-y(\tau))^2} {4(\tau-\tau_{j})} } e^{-\xi} U'_\xi(\tau_{j},\xi) d\xi \\
&= \int_0^\infty \frac{e^{-\frac{(Z + y(\tau_{j}) - y(\tau))^2} {4(\tau-\tau_{j})}}}{2\sqrt{\pi(\tau - \tau_{j})^3}} F_1(Z) e^{-(Z + y(\tau_{j}))}  U(\tau_{j},Z+y(\tau)) dZ, \nonumber \\
F_1(Z) &= \frac{\tau - \tau_{j}}{2} - 1 + \frac{(Z + \tau - \tau_{j} + y(\tau_{j}) - y(\tau))^2}{2(\tau-\tau_{j})}, \qquad Z = \xi - y(\tau_{j}), \nonumber
\end{align}
where, based on \eqref{Usol}
\begin{align} \label{Usol-1}
e^{- y(\tau_{j})} &e^{-Z} U(\tau_{j},Z+y(\tau_{j})) = \frac{1}{w_j} \int_{0}^{Z} \Phi(\tau_{j},k+y(\tau_{j})) e^{k} \exp \left( \frac{e^{k + y(\tau_{j})} - e^{Z}}{w_j} \right) dk, \\
\Phi(\tau,x) &= \int_0^\tau \int_{y(s)}^\infty e^{- s (\xi-y(s))^2} \frac{\eta(s,\xi)}{2\sqrt{\pi(\tau - s)}} \left[ e^{-\frac{(\xi - x)^2}{4(\tau - s)}} - e^{-\frac{(\xi - 2y(\tau) + x)^2}{4(\tau - s)}} \right] d\xi ds - w_j \zeta_\xi(\tau_{j},x), \nonumber \\
w_j &= \frac{2 \alpha(\tau_{j})}{\sigma^2(\tau_{j})} \frac{D_j}{K}. \nonumber
\end{align}

Since usually $D_j/K \ll 1$ and $\frac{2\alpha(\tau_{j})} {\sigma^2(\tau_{j})} = O(1)$, it follows that $w_j \ll 1$. Furthermore, the inequality $k + y(\tau_{j}) \le Z$ holds because $k \le Z$ and $y(\tau_{j}) < 0$. Consequently, the exponent in the integral in \eqref{Usol-1} is small everywhere except in a region near $Z=0$.
\begin{figure}[!htp]
\centering
\fbox{\includegraphics[width=0.7\textwidth]{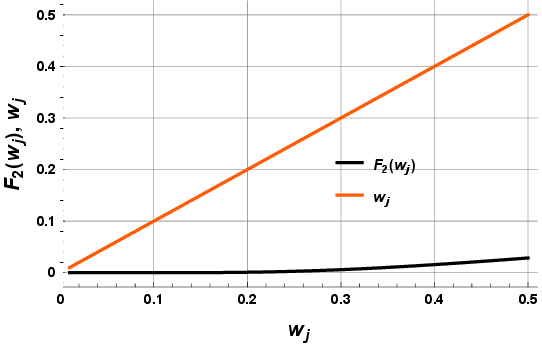}}
\caption{Comparison of $F_2(w_j)$ to the linear function $y(w_j) = w_j$.}
\label{F2wj}
\end{figure}

Given that $\Phi(\tau,Z+y(\tau_{j}))$ is bounded in $Z$, we apply Laplace's method, \cite{bender1978advanced} to obtain the leading-order asymptotics
\begin{align} \label{uPrimeInter}
e^{- (Z+y(\tau_{j}))} &U(\tau_{j},Z+y(\tau_{j})) \approx \frac{1}{w_j} e^{-\frac{e^{Z}}{w_j}} \lim_{Z \to 0} \int_0^Z \Phi(\tau,k+y(\tau_{j})) \exp \left(\frac{e^{y(\tau_{j}) + k}}{w_j} + k\right) \, dk \\
&= \Phi(\tau,y(\tau_{j})) \frac{1}{w_j} e^{\frac{e^{y(\tau_{j})}}{w_j}} Z e^{-\frac{e^{Z}}{w_j}} + O(Z^2). \nonumber
\end{align}

Given that $\tau > \tau_{j}$, the exponent in \eqref{J2Z} is bounded in $Z$. Substituting \eqref{uPrimeInter} into \eqref{J2Z}, therefore yields
\begin{align} \label{J2Z-1}
\calJ_1 &\approx \Phi(\tau,y(\tau_{j})) F_1(0) \frac{e^{ -\frac{( y(\tau_{j}) - y(\tau))^2}{4(\tau-\tau_{j})} + \frac{e^{y(\tau_{j})}}{w_j}}}{2\sqrt{\pi(\tau - \tau_{j})^3}} \frac{1}{w_j} \int_0^\infty Z e^{-\frac{e^{Z}}{w_j}} dZ \\
&= \Phi(\tau,y(\tau_{j})) F_1(0) \frac{e^{ -\frac{( y(\tau_{j}) - y(\tau))^2}{4(\tau-\tau_{j})} + \frac{e^{y(\tau_{j})}}{w_j}}}{2\sqrt{\pi(\tau - \tau_{j})^3}} F_2(w_j), \qquad
F_2(w_j) = \frac{1}{w_j} G_{2,3}^{3,0}\left(\frac{1}{w_j}|
\begin{array}{c}
 1,1 \\
 0,0,0 \\
\end{array}
\right), \nonumber
\end{align}
where the last $G$ symbol denotes the Meijer G-function, \cite{bateman1953higher}.

Fig.~\ref{F2wj} compares $F_2(w_j)$ to the linear function $y(w_j) = w_j$. The results show that $F_2(w_j)$ remains small even for unrealistically high values of $w_j$ near 50\%. Also, since $y(\tau_{j}) < 0$, the exponent in \eqref{J2Z-1} is small. Therefore, the integral $\calJ_1$ is negligible and can be disregarded in a good approximation.

\end{document}